\documentclass[12pt]{article}
\usepackage{putex}
\usepackage{graphicx}
\usepackage{epstopdf}
\usepackage{enumerate}
\usepackage{psfrag}
\usepackage{subfigure}
\usepackage{cite}

\numberwithin{equation}{section}

\newcommand{\lapten}{\square}
\newcommand{\lap}{\Delta}
\newcommand{\grp}[1]{\mathrm{#1}}

\newcommand{\grO}{\grp{O}}
\newcommand{\grU}{\grp{U}}
\newcommand{\grSU}{\grp{SU}}
\newcommand{\grSO}{\grp{SO}}

\newcommand{\grGL}{\grp{GL}}

\newcommand{\abs}[1]{\lvert #1 \rvert}

\newcommand {\bes} {\begin {equation*}}
\newcommand {\ees} {\end {equation*}}

\newcommand{\es}[2] {\begin{equation} \label{#1} \begin{split} #2 \end{split} \end{equation}}

\newcommand{\Z}{\mathbb{Z}}
\newcommand{\R}{\mathbb{R}}
\newcommand{\C}{\mathbb{C}}

\def\SSP#1{{\Red [SSP: #1]}}
\def\TK#1{{\Green [TK: #1]}}
\def\JL#1{{\Blue [JL: #1]}}

\def\SSP#1{}
\def\TK#1{}
\def\JL#1{}

\renewcommand{\Re}{\mathop{\mathrm{Re}}}
\renewcommand{\Im}{\mathop{\mathrm{Im}}}

\newcommand{\order}{\mathcal{O}}

\newcommand{\matr}[2]{\left(\begin{array}{#1}#2\end{array}\right)}
\newcommand{\rep}[1]{{\mathbf{#1}}}

\ifx\genfrac\sdflkaj

\else

\fi
\newcommand{\sfrac}[2]{{\textstyle\frac{#1}{#2}}}
\newcommand{\half}{\sfrac{1}{2}}

\newcommand{\quarter}{\sfrac{1}{4}}

\newcommand{\Half}{\frac{1}{2}}

\newcommand{\Quarter}{\frac{1}{4}}

\newcommand{\brk}[1]{(#1)}
\newcommand{\lrbrk}[1]{\left(#1\right)}
\newcommand{\bigbrk}[1]{\bigl(#1\bigr)}
\newcommand{\Bigbrk}[1]{\Bigl(#1\Bigr)}
\newcommand{\biggbrk}[1]{\biggl(#1\biggr)}

\newcommand{\lrsbrk}[1]{\left[#1\right]}
\newcommand{\bigsbrk}[1]{\bigl[#1\bigr]}
\newcommand{\Bigsbrk}[1]{\Bigl[#1\Bigr]}
\newcommand{\biggsbrk}[1]{\biggl[#1\biggr]}
\newcommand{\Biggsbrk}[1]{\Biggl[#1\Biggr]}
\newcommand{\ket}[1]{\mathopen{|}#1\mathclose{\rangle}}

\newcommand{\bigabs}[1]{\bigl|#1\bigr|}

\newcommand{\bigfloor}[1]{\bigl\lfloor#1\bigr\rfloor}

\newcommand{\lrfloor}[1]{\left\lfloor#1\right\rfloor}
\newcommand{\bigceiling}[1]{\bigl\lceil#1\bigr\rceil}

\newcommand{\be}{\begin{eqnarray}}
\newcommand{\ee}{\end{eqnarray}}
\newcommand{\nn}{\nonumber}
\newcommand{\nln}{\nonumber\\}
\newcommand{\nl}[1][0pt]{\nonumber\\[#1]&\hspace{-4\arraycolsep}&\mathord{}}

\newcommand{\earel}[1]{\mathrel{}&\hspace{-2\arraycolsep}#1\hspace{-2\arraycolsep}&\mathrel{}}
\newcommand{\eq}{\earel{=}}

\makeatletter
\def\mr@ignsp#1 {\ifx\:#1\@empty\else #1\expandafter\mr@ignsp\fi}%
\newcommand{\multiref}[1]{\begingroup
\xdef\mr@no@sparg{\expandafter\mr@ignsp#1 \: }%
\def\mr@comma{}%
\@for\mr@refs:=\mr@no@sparg\do{\mr@comma\def\mr@comma{,}\ref{\mr@refs}}%
\endgroup}
\makeatother

\newcommand{\hypref}[2]{\ifx\href\asklfhas #2\else\href{#1}{#2}\fi}

\renewcommand{\eqref}[1]{(\multiref{#1})}

\newcommand{\eps}{\varepsilon}

\DeclareMathOperator{\arccosh}{arccosh}

\newcommand{\by}{\bar{y}}

\begin{document}

\preprint{PUPT-2349\\ UUITP-30/10}

\institution{PU}{$^1$Joseph Henry Laboratories and $^2$Center for Theoretical Science,\cr
~~~~~~~~~~~~~~Princeton University, Princeton, NJ 08544}

\institution{UU}{$^3$Department of Physics and Astronomy, Uppsala University,\cr
~~~~~~~~~~~~~~~~~~~~~~~~~SE-75120 Uppsala, Sweden}

\institution{UC}{$^4$Department of Physics, University of Chicago, Chicago, IL 60637}

\title{Green's Functions and Non-Singlet Glueballs on Deformed Conifolds}

\authors{Silviu S.~Pufu,$^1$ Igor R.~Klebanov,$^{1,2}$ Thomas Klose,$^3$ and Jennifer Lin$^{1, 4}$ }

\abstract{
We study the Laplacian on Stenzel spaces (generalized deformed conifolds), which are tangent bundles of spheres endowed with Ricci flat metrics.  The $(2d-2)$-dimensional Stenzel space has $\grSO(d)$ symmetry and can be embedded in $\C^d$ through the equation $\sum_{i = 1}^d {z_i^2} = \epsilon^2$.  We discuss the Green's function with a source at a point on the $S^{d-1}$ zero section of $TS^{d-1}$.  Its calculation is complicated by mixing between different harmonics with the same $\grSO(d)$ quantum numbers due to the explicit breaking by the $\epsilon$-deformation of the $\grU(1)$ symmetry that rotates $z_i$ by a phase.  A similar mixing affects the spectrum of normal modes of warped deformed conifolds that appear in gauge/gravity duality. We solve the mixing problem numerically to determine certain bound state spectra in various representations of $\grSO(d)$ for the $d=4$ and $d=5$ examples.
}

\date{September 2010}

\maketitle

\tableofcontents

\section{Introduction}

An interesting class of generalizations of the AdS/CFT duality \cite{Maldacena:1997re, Gubser:1998bc, Witten:1998qj} involves theories which exhibit color confinement and discrete spectra of bound states \cite{Witten:1998zw, Klebanov:2000hb}.
In such a theory, the bound state masses $m_i$ can be read off from the poles of two-point functions of operators in Fourier space,
 \es{TwoPoint}{
  \langle {\cal O}(k) {\cal O}(-k) \rangle
   \sim \sum_i \frac{c_i}{k^2 + m_i^2} + \text{less singular terms} \, .
 }
In gauge/gravity duality, such poles correspond to normalizable solutions of the linearized supergravity equations of motion for the bulk field $\Phi$ dual to the operator ${\cal O}$.

The first studies of bound state (glueball) spectra via gauge/gravity duality \cite{Csaki:1998qr, deMelloKoch:1998qs, Brower:2000rp} were carried out for non-supersymmetric backgrounds proposed in \cite{Witten:1998zw} that describe confinement in $(3+1)$ and in $(2+1)$ dimensions. These backgrounds have $\grSO(5)$ ($\grSO(6)$) symmetry due to the presence of an $S^4$ ($S^5$) in the geometry.  The glueballs that are non-singlet under these global symmetries were investigated early on \cite{Ooguri:1998hq}, with the result that their masses are comparable to those of the singlet glueballs.

The methods of gauge/gravity duality were subsequently applied to calculation of bound state spectra in the cascading $\grSU((k+1)M)\times \grSU(kM)$ supersymmetric gauge theory coupled to bifundamental chiral superfields $A_1, A_2, B_1, B_2$ \cite{Klebanov:2000nc,Klebanov:2000hb}. This theory has a global $\grSO(4)\cong \grSU(2)_L\times \grSU(2)_R$ flavor symmetry that acts on the chiral superfields. Therefore, it is expected that the bound state spectra form Kaluza-Klein towers with increasing $\grSO(4)$ quantum numbers.  The gravity dual of the cascading gauge theory is the warped deformed conifold (KS) background \cite{Klebanov:2000hb}, which is a warped product of $\R^{3, 1}$ and the deformed conifold \cite{Candelas:1989js} described by the locus $\sum_{i = 1}^4 z_i^2 = \epsilon^2$ in $\C^4$. This space has an obvious $\grSO(4)$ symmetry that acts on the $z$-coordinates. Therefore, the normal modes of this space can be classified according to their $\grSO(4)$ quantum numbers. The study of glueball masses in the $\grSO(4)$-invariant sector was initiated in \cite{Krasnitz:2000ir} and continued in \cite{Gubser:2004qj,Berg:2005pd,Berg:2006xy, Dymarsky:2007zs,Benna:2007mb, Dymarsky:2008wd,Gordeli:2009nw}.  One of the goals of this paper is to initiate a systematic study of glueballs that are not $\grSO(4)$ singlets (see also \cite{Firouzjahi:2005qs}).

We will focus on the case of a minimally coupled scalar field in the bulk, namely a field $\Phi$ that satisfies the 10-d Laplace equation
 \es{MinimallyCoupled}{
  \lapten \Phi = 0 \; .
 }
Such an equation of motion arises, for example, from transverse metric fluctuations in the $\R^{3, 1}$ part of the KS geometry.
The ten-dimensional type IIB metric of the KS background is \cite{Klebanov:2000hb,Herzog:2001xk}
 \es{typeIIBMetric}{
  ds^2 = H^{-{1 \over 2}} dx_\mu dx^\mu + H^{1 \over 2} ds_6^2 \; ,
 }
where $dx_\mu dx^\mu$ is the Minkowski metric on $\R^{3, 1}$, $ds_6^2$ is the Calabi-Yau metric on the deformed conifold, and the warp factor $H$ is an $\grSO(4)$-invariant function on the deformed conifold.  Translational invariance in the $\R^{3, 1}$ directions allows us to look for plane wave solutions to eq.~\eqref{MinimallyCoupled} of the form $\Phi = e^{i k \cdot x} \phi$, where $\phi$ depends only on the six internal coordinates.  With this ansatz, the ten-dimensional Laplace equation takes the form of the eigenvalue problem
   \es{EValue}{
    \lap_6 \phi = -m^2 H \phi \; ,
   }
where $m^2 = -k_\mu k^\mu$ and $\lap_6$ is the Laplacian on the deformed conifold.

When $\Phi$ is an $\grSO(4)$ singlet, the dual operator in the boundary theory is the stress-energy tensor.  The higher $\grSO(4)$ harmonics of $\Phi$ correspond to single trace operators which are roughly the stress-energy tensor multiplied by polynomials in the bi-fundamental superfields of the KS theory and their Hermitian conjugates. The deformation $\epsilon$ breaks the $\grU(1)_R$ symmetry of the conifold CFT \cite{Klebanov:1998hh} which rotates all the chiral superfields by the same phase (in the cascading gauge theory, the $\grU(1)_R$ is broken to $\Z_{2M}$ by the chiral anomaly \cite{Klebanov:2002gr}, but for large $M$ this group may be viewed as a discrete ``approximation'' to $\grU(1)_R$). Therefore, operators that have the same $\grSO(4)$ quantum numbers but different $\grU(1)_R$ charges mix in the glueball calculation. This infrared mixing of modes in general necessitates solving coupled radial equations whose number typically grows with the size of the $\grSO(4)$ representation. This feature adds a new interesting structure to the non-singlet glueball calculations that was not encountered in \cite{Ooguri:1998hq}.

A problem closely related to the Kaluza-Klein spectrum of glueballs is the calculation of the Green's function on the deformed conifold.  The Green's function is of interest in gauge/gravity duality because it determines the extra term in the warp factor produced by a stack of $p$ D3-branes on the deformed conifold \cite{Krishnan:2008gx}. The deformed conifold with $p$ added mobile D3-branes provides a string dual of the cascading $\grSU((k+1)M+p)\times \grSU(kM+p)$ gauge theory \cite{Dymarsky:2005xt}. As usual, the Green's function may be expanded in an infinite set of eigenstates of the Laplacian. We find, though, that the calculation is considerably more complicated than in the case of the resolved conifold \cite{Klebanov:2007us}. For the deformed conifold, finding the eigenstates requires solving the same kind of coupled differential equations that appear in the non-singlet glueball calculations, but with different boundary conditions at the tip.

We will also consider analogous calculations in the weakly curved M-theory background found by Cvetic, Gibbons, Lu, and Pope (CGLP) \cite{Cvetic:2000db}.  This background is a warped product of $\R^{2, 1}$ and the $\grSO(5)$ symmetric Stenzel space \cite{Stenzel}, $\sum_{i = 1}^5 {z_i^2} = \epsilon^2$, which is a higher-dimensional generalization of the deformed conifold. The CGLP background \cite{Cvetic:2000db} is similar to the KS solution \cite{Klebanov:2000hb}, but it is asymptotic to $AdS_4\times V_{5,2}$ without any UV logarithms. In the infrared the background contains a blown-up 4-sphere, and the warp factor approaches a finite value.
Some aspects of the infrared physics were discussed in \cite{Herzog:2000rz,Martelli:2009ga,Jafferis:2009th}, but the dual infrared gauge theory remains to be elucidated. The CGLP background has a discrete spectrum of normal modes which describe bound states in the dual field theory \cite{JLthesis}. They may be classified according to their $\grSO(5)$ quantum numbers, and we will present the bound state spectra for a few representations.

The rest of this paper is organized as follows.  In section~\ref{GenerDef}, we explain how one can parameterize the Stenzel spaces \cite{Stenzel}, $\sum_{i = 1}^d z_i^2 = \epsilon^2$, in terms of a radial coordinate $\tau$ and the coordinates $y_i$ of the undeformed cone with $\epsilon =0$. The base of this cone is the Stiefel manifold $V_{d,2} = \grSO(d) / \grSO(d-2)$, on which we define a basis of normalizable functions. 
In section~\ref{LAPLACIAN}, we focus on $d=4$ corresponding to the $6$-dimensional deformed conifold. We use the $\grSO(4)$ symmetry to write the Laplacian $\lap_6$ explicitly. In section~\ref{GREEN}, we discuss the calculation of the Green's function of this Laplacian with a source placed on the blown-up 3-sphere. In section~\ref{DECOUPLING} we present a complete set of basis functions on the deformed conifold. In section~\ref{GLUEBALLSIIB}, we use this basis to derive the eigenvalue problem for the non-singlet glueballs. We present our numerical results for the glueball masses belonging to the first few $\grSO(4)$ representations, and compare them with the WKB analysis.  In section~\ref{GENERALIZATION}, we generalize our results for the Laplacian and Green's function to Stenzel spaces with $d>4$. In section~\ref{GLUEBALLSM}, we compute numerically the glueball masses in the theory dual to the M-theory background of \cite{Cvetic:2000db}.
We end the paper with a discussion of our results and of some open problems.


\section{Generalized deformed conifolds}
\label{GenerDef}

\subsection{Coordinates and metrics}
\label{PARAMETERIZATION}

The $(2d - 2)$-dimensional Stenzel space \cite{Stenzel} is a Calabi-Yau manifold defined by the quadric
\es{DeformedConifold}{
  \sum_{i = 1}^{d} z_i^2 = \epsilon^2
}
in $\C^{d}$, where the constant $\epsilon$ can be taken to be real and positive. For $d=3$ this space is the Eguchi-Hanson space \cite{Eguchi:1978xp}, while for $d=4$ it is the deformed conifold \cite{Candelas:1989js}. We will sometimes refer to Stenzel spaces with arbitrary $d$ as a ``generalized deformed conifold.''

When the deformation parameter $\epsilon$ vanishes, the space described by eq.~\eqref{DeformedConifold} is a cone because its definition becomes invariant under rescalings $z_i \to t z_i$.  The base of this undeformed cone is given by
 \es{yConstraints}{
  \sum_{i = 1}^{d} y_i^2 = 0\; , \qquad \sum_{i = 1}^d \abs{y_i}^2 = 1 \; ,
 }
where in order to avoid confusion with the deformed conifold \eqref{DeformedConifold}, we denote the complex coordinates in $\C^d$ by $y_i$ instead of $z_i$.  The space in \eqref{yConstraints} is also known as the $(2d-3)$-dimensional Stiefel manifold $V_{d, 2}$ defined as the set of pairs of orthonormal vectors $\vec{u}$ and $\vec{v}$ in $\R^d$,
 \es{onVectors}{
  \abs{\vec{u}}^2 = \abs{\vec{v}}^2 = 1 \; , \qquad
   \vec{u} \cdot \vec{v} = 0 \; .
 }
These vectors parameterize the coset space $\grSO(d) / \grSO(d-2)$.  That \eqref{yConstraints} and \eqref{onVectors} are the same can be seen by the explicit embedding
 \es{yDef}{
  y_i = {1\over \sqrt{2}} \left( u_i + i v_i \right)
 }
of the Stiefel manifold in $\C^d \cong \R^d \times \R^d$.

The Stenzel space \eqref{DeformedConifold} can then be parameterized by the $y_i$ and a ``radial'' coordinate $\tau$:
 \es{zToy}{
  z_i = {\epsilon \over \sqrt{2}} \left(e^{\tau \over 2} y_i + e^{-{\tau \over 2}} \bar y_i \right)  \; .
 }
The constraints \eqref{yConstraints} guarantee that the defining relation \eqref{DeformedConifold} is obeyed.  In order to cover the Stenzel space only once, the range of $\tau$ should be either $0 \leq \tau < \infty$ or $-\infty <\tau \leq 0$.  These two choices are equivalent, so in the rest of this paper we will focus on $0 \leq \tau < \infty$.  As will be developed more fully in section~\ref{TOPOLOGY}, the constant $\tau>0$ surfaces in the Stenzel space are much like $V_{d, 2}$:  they have the same topology as $V_{d, 2}$, so on them one can define the same functions as on $V_{d, 2}$.

As on any K\"ahler manifold, the metric on this generalized deformed conifold can be written in terms of a K\"ahler potential ${\cal F}$ as  \es{Metric}{
  ds_{2 d - 2}^2 = g_{i\bar{\jmath}} \, dz_i d\bar{z}_j \; ,
  \qquad
  g_{i\bar{\jmath}} = \frac{\partial^2 {\cal F}}{\partial z_i \partial \bar z_j}  \; .
 }
We are only interested in Ricci flat metrics that obey an $\grO(d) \cong \grSO(d)\times \Z_2$ symmetry, where the $\grSO(d)$ rotates the $z_i$ into one another\footnote{Because of the $\grSO(d)$ symmetry (or rather the absence of an $\grSU(d)$ symmetry) we make no distinction between upper and lower indices on the $y_i$ and $z_i$ coordinates.} and the $\Z_2$ acts by flipping the sign of one of the $z_i$. The $\grSO(d)$ invariance constrains  $\mathcal{F}$ to be a function of $\tau$, so that
\es{MetricAgain}{
  g_{i\bar{\jmath}} = \frac{\mathcal{F}'(\tau)}{\epsilon^2 \sinh\tau} \, \delta_{i\bar{\jmath}}
                    + \frac{\mathcal{F}''(\tau)-\mathcal{F'}(\tau)\coth\tau }{\epsilon^4 \sinh^2\tau} \, \bar{z}_i z_j \; ,
}
where the primes denote differentiation with respect to $\tau$.  The Ricci flatness condition reduces to a first order differential equation for $\mathcal{F}'(\tau)$ given by \cite{Cvetic:2000db}\footnote{For the relation between our conventions and the ones in \cite{Cvetic:2000db}, see Appendix~\ref{V52METRIC}.  See also section~\ref{GENERALIZATION} below for more details on how to obtain \eqref{eqn:ricciflat}.}
\be \label{eqn:ricciflat}
  \lrbrk{\frac{\mathcal{F}'}{\epsilon^2 \sinh\tau}}^{d-2} \mathcal{F}'' = \frac{d-2}{d-1} \; .
\ee
The solution can be written as
\be
  \mathcal{F}' = \epsilon^2 R(\tau)^{\frac{1}{d-1}}
  \qquad\text{with}\qquad
  R(\tau) \equiv {d-2 \over \epsilon^2} \int_0^\tau (\sinh v)^{d-2} \, dv \; .
\ee

Since the Stenzel metric \eqref{Metric} has an $\grSO(d)$ symmetry, it should admit $d(d-1)/2$ Killing vectors.  They are
 \es{KillingVectors}{
  \xi^{(a)} = T^{(a)}_{ij} y_i \frac\partial{\partial y_j} + T^{(a)}_{ij} \bar y_i \frac\partial{\partial \bar y_j}  \; ,
 }
where the index $a$ runs from $1$ to $d(d-1)/2$ and the $T^{(a)}_{ij}$, which are real $d \times d$ antisymmetric matrices, are the generators of $\grSO(d)$.  One of the reasons that the parameterization \eqref{zToy} is convenient is that the Killing vectors \eqref{KillingVectors} have simple expressions in terms of the $y_i$. In these coordinates, the metric \eqref{MetricAgain} reads
\be\label{dsy}
  ds_{2d-2}^2 \eq \quarter \mathcal{F}'' d\tau^2
       + \mathcal{F}' \coth\tau \, dy_i d\bar{y}_i \nl
       + \half \mathcal{F}' \csch\tau \, \bigbrk{ dy_i dy_i + d\bar{y}_i d\bar{y}_i }
       + \bigbrk{ \mathcal{F}'' - \mathcal{F}' \coth\tau } \, y_i d\bar{y}_i \bar{y}_j dy_j \; .
\ee

\subsection{Topology and group actions}
\label{TOPOLOGY}

The meaning of the coordinate $\tau$ and the relevance of $V_{d, 2}$ in parameterizing the generalized deformed conifolds can be uncovered by examining the topology of these spaces.  One can start by writing the coordinates $z_i$ in terms of the orthonormal vectors $u_i$ and $v_i$ in $\R^d$:
 \es{zTouv}{
  z_i = \epsilon \left( u_i \cosh {\tau \over 2} + i v_i \sinh {\tau \over 2} \right) \; .
 }
If we are interested only in the topology of the generalized deformed conifold parameterized by the $z_i$, we can deform this space smoothly to the set of points in $\C^n$ given by
 \es{DifferentSpace}{
  u_i + i \tau v_i \; ,
 }
where the range of $\tau$ is still $0 \leq \tau < \infty$.  The real part of this expression parameterizes a unit $S^{d-1} \subset \R^d$, while the imaginary part parameterizes those vectors in $\R^d$ that are tangent to it, showing that the Stenzel space \eqref{DeformedConifold} has the topology of the tangent bundle $TS^{d-1}$.\footnote{The symplectic structure induced from $\C^d$ agrees with the standard
symplectic structure on the cotangent bundle, $T^*S^{d-1}$ \cite{Cvetic:2000db}.  The cotangent bundle $T^* S^{d-1}$ and the tangent bundle $TS^{d-1}$ are homeomorphic as $2(d-1)$-manifolds, so there is no topological distinction between them.}  Up to reparameterizations, the coordinate $\tau$ can be identified with the radial coordinate of the $\R^{d-1}$ fiber of $TS^{d-1}$.  The $\tau = 0$ section of the tangent bundle, $S^{d-1}$, is the only non-trivial cycle of the generalized deformed conifold.  For $\tau>0$, each constant $\tau$ section is described by pairs of orthogonal vectors of fixed length in $\R^d$, so it is homeomorphic to $V_{d, 2}$.  Indeed, $V_{d, 2}$ is topologically the $S^{d-2}$ fiber bundle over $S^{d-1}$ obtained by restricting the $\R^{d-1}$ fiber in $TS^{d-1}$ to the unit $S^{d-2} \subset \R^{d-1}$.   This bundle is trivial if and only if $d = 2, 4$, or $8$ \cite{Kervaire, MR0102804}, but it has the homology of a product of spheres, $S^{d-2} \times S^{d-1}$, if and only if $d$ is even \cite{Hatcher}.

We can also understand the topology of a generalized deformed conifold by examining the action of $\grSO(d)$ on this space.  The group $\grSO(d)$ acts on the Stenzel space as isometries by rotating the complex coordinates $z_i$ or $y_i$ into each other, while leaving $\tau$ invariant.  This group action is transitive on any constant $\tau$ surface $\Sigma_\tau$, as any point on that surface can be obtained by acting with an $\grSO(d)$ matrix $O$ on the point
 \es{SpecialPoint}{
  \vec{z}_* = \epsilon \begin{pmatrix} 0 & 0 & \cdots & 0 & i \sinh {\tau \over 2} & \cosh {\tau \over 2} \end{pmatrix} \; .
 }
When $\tau>0$, the only redundancy in this description is that the matrix $O$ can be multiplied on the right by any rotation matrix in the first $d-2$ coordinates without changing $O \vec{z}_*$.  So every $\Sigma_\tau$ with $\tau>0$  can be identified with the space $\grSO(d) / \grSO(d-2) = V_{d, 2}$.  When $\tau = 0$, the redundancy is an $\grSO(d-1)$ subgroup of $\grSO(d)$ corresponding to rotations in the first $d-1$ coordinates, so $\Sigma_0$ has the topology of $\grSO(d) / \grSO(d-1) = S^{d-1}$.

\subsection{Normalizable functions on Stiefel manifolds}
\label{FUNCTIONS}

The two problems we are going to address in this paper, namely that of finding the Green's function of the Laplacian on the Stenzel space with a source on the $S^{d-1}$ at $\tau=0$, and that of finding the eigenfunctions of the Laplacian, involve dealing with functions that are smooth at any point on $\Sigma_\tau$ with $\tau>0$.  The strategy that we will adopt is to use group theory to find a basis of normalizable functions on each $\Sigma_\tau$ surface, and then expand the functions on the Stenzel space in terms of this basis.

Let us focus on the sections $\Sigma_\tau$ with $\tau>0$.  While it is certainly true that there is some freedom in what the induced metric on $\Sigma_\tau$ could be,\footnote{The most general $\grSO(d)$-invariant metric is a linear combination of $\Re (d\vec{y} \cdot d\vec{y})$, $\Im (d\vec{y} \cdot d\vec{y})$, $\abs{d\vec{y}}^2$, and $\abs{\vec{\bar y} \cdot d\vec{y}}^2$ with arbitrary, possibly $\tau$-dependent, coefficients.} the integration measure on this space is fixed by the $\grSO(d)$ symmetry up to an overall $\tau$-dependent normalization.  In fact, if $\mu_{\grSO(d)}$ is the unit-normalized Haar measure on $\grSO(d)$, a canonical way of finding a measure $\mu_{\grSO(d)/\grSO(d-2)}$ on the coset space $\grSO(d) / \grSO(d-2)$ is through the pushforward operation, where each measurable set in $\grSO(d) / \grSO(d-2)$ is assigned a measure equal to the measure of its preimage under the projection map $\grSO(d) \to \grSO(d) / \grSO(d-2)$.  Therefore, up to an overall $\tau$-dependent normalization, the measure on $\Sigma_\tau$ is given by $\mu_{\grSO(d)/\grSO(d-2)}$.  As a consequence, the space $L^2(\Sigma_\tau)$ of square integrable functions on $\Sigma_\tau$ is the same for all $\tau$, being equal to the space $L^2(V_{d, 2})$ with the integration measure $\mu_{\grSO(d) / \grSO(d-2)}$.

The square integrable functions $L^2(V_{d, 2})$ were studied in \cite{Strichartz, Gelbart1, Gelbart2, Levine}, so most of the results presented in the rest of this section are not new.  As a vector space acted on by $\grSO(d)$, $L^2(V_{d, 2})$ decomposes into irreps of $\grSO(d)$.  This decomposition can be understood from group theory as follows.  Any function on $V_{d, 2} = \grSO(d) / \grSO(d-2)$ can be thought of as a function on $\grSO(d)$ that is constant on each coset, namely $f(g) = f(gh)$, where $g$ is an arbitrary element of $\grSO(d)$ and $h$ is an arbitrary element of $\grSO(d-2) \subset \grSO(d)$.  In order to understand the normalizable functions on $V_{d, 2}$ we should first understand the normalizable functions on $\grSO(d)$ itself.  To construct functions on $\grSO(d)$, consider an irreducible representation of $\grSO(d)$ on a Hilbert space $V$ with an orthonormal basis $e_k$, $1 \leq k \leq \dim V$.  The matrix element $M_{ij}^{V}(g) = \langle e_i | g | e_j \rangle$ is naturally a function on $\grSO(d)$, and by construction it is a polynomial in the entries of $g$ whose degree equals the number of boxes in the Young diagram corresponding to $V$.  There is a theorem due to Peter and Weyl (see for example \cite{Taylor}) that asserts that the set of all such matrix elements between basis states for all irreducible representations of a compact Lie group, in particular $\grSO(d)$, forms a basis for the set of square integrable functions on that group, in our case $L^2(\grSO(d))$.  Under the left-action of $\grSO(d)$ on $L^2 (\grSO(d))$ (defined as $\tilde g f(g) = f(\tilde g^{-1} g)$), each column in $M_{ij}^V(g)$ transforms like the basis of $V$, and under the right action of $\grSO(d)$ each row transforms like the dual basis. \SSP{Check previous statement.}  A consequence of the Peter-Weyl theorem is that under the decomposition of $L^2(\grSO(d))$ under the left action of $\grSO(d)$, each irrep $V$ appears with multiplicity $\dim V$, because in the matrix of basis functions $M_{ij}^V(g)$ there are $\dim V$ columns that transform in exactly the same way.

In constructing a basis of functions for $L^2(V_{d, 2})$, we want to retain only those linear combinations of $M_{ij}^V(g)$ that are invariant under the right action of $\grSO(d-2)$.  Since $\grSO(d-2)$ acts only within each row of $M_{ij}^V(g)$, from each row we should keep only those linear combinations of basis functions that are invariant under $\grSO(d-2)$.  It follows that in the decomposition of $L^2(V_{d, 2})$ under the left action of $\grSO(d)$, each representation $V$ appears a number of times equal to the number of singlets in the decomposition of $V$ under $\grSO(d-2)$.

As an example, let us discuss briefly the case $d = 4$.   The five-dimensional Stiefel manifold $V_{4, 2} = \grSO(4) / \grSO(2)$ is commonly referred to as $T^{1, 1}$ \cite{Romans:1984an}.  The group $\grSO(2)$ representing rotations in the first two coordinates of $\vec{z}_*$ commutes only with rotations in the last two coordinates of $\vec{z}_*$, so it can be chosen to be generated by $J_{L3} - J_{R3}$, where $\vec{J}_L$ and $\vec{J}_R$ are the generators of the left and right $\grSU(2)$ factors in $\grSO(4)\cong \grSU(2)_L \times \grSU(2)_R$.  For the spin $[j_L, j_R]$ irrep, which has a total of $(2 j_L + 1)(2j_R+1)$ basis states, the number of $\grSO(2)$ singlets equals the number of basis states with $m_L = m_R$, where $m_L$ and $m_R$ are the left and right magnetic quantum numbers.  Since $m_L$ ($m_R$) ranges between $\pm j_L$ ($\pm j_R$), the number of such states is $2 \min(j_L, j_R) + 1$.    The discussion in the previous paragraph shows that on $T^{1, 1}$ there must be precisely $2 \min(j_L, j_R) + 1$ normalizable functions with the same $\grSO(4)$ quantum numbers.  Indeed, the fundamental representation of $\grSO(4)$, $[\half,\half]$, is represented once by $y_i$ and once by $\bar y_i$;  the $[1, 1]$ irrep appears three times:  $y_i y_j$, $\bar y_i \bar y_j$, and $y_i \bar y_j - {1 \over 4} \delta_{ij}$, etc.

In general, to construct explicitly a basis for $L^2(V_{d, 2})$, consider polynomials in $y_i$ and $\bar y_i$
 \es{Polyn}{
  F(y_i, \bar y_i) = M_{i_1 i_2 \cdots i_{n_1}}^{j_1 j_2 \cdots j_{n_2}} \,
    y_{i_1} y_{i_2} \cdots y_{i_{n_1}} \, \bar y_{j_1} \bar y_{j_2} \cdots \bar y_{j_{n_2}} \; ,
 }
where $M_{i_1 i_2 \cdots i_{n_1}}^{j_1 j_2 \cdots j_{n_2}}$ are constant, linearly independent $\grSO(d)$ tensors that are symmetric in the lower and upper indices separately.  A choice of basis of square normalizable functions on $V_{d, 2}$ consists of the basis of polynomials in eq.~\eqref{Polyn} modulo the constraints imposed by eq.~\eqref{yConstraints}.  In constructing this basis, it is convenient to choose the linearly-independent tensors $M_{i_1 i_2 \cdots i_{n_1}}^{j_1 j_2 \cdots j_{n_2}}$ to transform irreducibly under $\grSO(d)$. These basis elements can be thought of as those functions $M_{ij}^V(g)$---which, as discussed above, are polynomials in the entries of the $\grSO(d)$ matrix $g$---that are invariant under the right action of $\grSO(d-2)$.

As can be seen from the $\grSO(4)$ example presented above, for a given irrep $V$ of $\grSO(d)$, the linearly-independent polynomials \eqref{Polyn} with the same $\grSO(d)$ quantum numbers can be distinguished by the total number of $y_i$ minus the total number of $\bar y_i$.  Later on we will call this quantity $\tilde m$.  In terms of the undeformed conifold, $\tilde m$ is identified with half the R-charge of the dual operator.  Thinking of $y_i$ and $\bar y_i$ as independent complex variables, a more detailed analysis \cite{Gelbart1, Gelbart2} reveals that the polynomials with the same $\grSO(d)$ quantum numbers in $V$ also transform in an irrep of $\grGL(2, \C)$ for which $\tilde m$ is the quantum number of a $\grU(1)$ subgroup of $\grGL(2, \C)$.  We will provide more details in section~\ref{DECOUPLING}.

To construct wavefunctions $\phi(\tau, y_i, \bar y_i)$ on the deformed conifold, we just take linear combinations of polynomials \eqref{Polyn} with coefficients dependent on $\tau$:
 \es{GeneralWavefunction}{
  \phi(\tau, y_i, \bar y_i) = \sum_\alpha f_\alpha(\tau) F_\alpha (y_i, \bar y_i) \; ,
 }
where $F_\alpha (y_i, \bar y_i)$ is a state in a definite $\grSO(d)$ representation.

\subsection{Decoupling of differential equations}

The decomposition \eqref{GeneralWavefunction} is quite powerful when solving for the Green's function corresponding to the Laplacian or when solving an eigenvalue problem like in eq.~\eqref{EValue}.  The Laplace operator on the Stenzel space $\lap_{2d-2}$ is $\grSO(d)$-invariant, so it can only mix terms in \eqref{GeneralWavefunction} with the same $\grSO(d)$ quantum numbers.  Therefore, in order to find the coefficient functions $f_\alpha(\tau)$ for a given $\grSO(d)$ irrep $V$, we only need to solve a coupled system of ODEs. The number of coupled equations is at most the number of linearly independent functions on $V_{d, 2} = \grSO(d) / \grSO(d-2)$ with the same $\grSO(d)$ quantum numbers.  As discussed above, the number of such functions equals the number of singlets in the decomposition of $V$ under $\grSO(d-2)$.

Note that there is no conserved R-charge in the case of the deformed conifold. The would-be $\grU(1)_R$-symmetry that rotates the $z$'s by a phase is explicitly broken by the non-zero $\epsilon$ in \eqref{DeformedConifold}. Nor is there an R-symmetry that rotates the $y$'s by a phase, as can be seen from the third term in the metric \eqref{dsy}. Only in the limit of large $\tau$, where the $\epsilon$-deformation becomes insignificant and the $z$'s and $y$'s are essentially the same, can one define a conserved R-charge. Thus, although asymptotically the equations do not couple functions with a different number of $y$'s minus $\bar{y}$'s, there will in general be mixing. Even for finite $\tau$ however, there is a $\Z_2$ symmetry that exchanges $y_i \leftrightarrow \bar y_i$.  Since it interchanges chiral and anti-chiral operators we identify this symmetry with parity.  Its existence leads to a decoupling of the functions that are even and odd under it. Hence, for a generic $\grSO(d)$ representation, one has to solve two sets of coupled ODEs.

As a side comment, we note that group theory can be used in similar ways to simplify the finding of the Green's function or of the eigenfunctions of the Laplacian on more general spaces.  The manifold $V_{d, 2}$ in the above discussion could be replaced by a manifold $X$ on which an isometry group $G$ acts with cohomogeneity $k$ (i.e.~for which the generic orbits of $G$ have codimension $k$).   The space $X$ can be foliated by orbits of points in $X$ under $G$.  If $K$ is the stabilizer of a generic point, then the orbit of this point has the topology of the space $G/K$.  That $G$ acts on $X$ as an isometry means that each such generic leaf inherits from $X$ a measure proportional to the pushforward of the unit-normalized Haar measure on $G$, the coefficient of proportionality being allowed to vary from leaf to leaf as specified by a function of the $k$ ``transverse'' ($\tau$-like) coordinates.  The space of $L^2$-normalizable functions on each generic leaf, $L^2(G/K)$, is an infinite-dimensional vector space acted on by $G$, so it decomposes into irreps of $G$.  As a consequence of the Peter-Weyl theorem, the multiplicity of a given irrep $V$ in this decomposition equals the number of singlets in the decomposition of $V$ under $K$.  In finding the Green's function or the eigenfunctions of the Laplacian, one has to solve in this case a system of PDEs in the $k$ transverse coordinates for each irrep $V$ of $G$.  The number of equations in each set equals the number of singlets in the decomposition of $V$ under $K$.   Clearly, the number of coupled equations is at most equal the dimension of $V$, and equality holds when each generic leaf in the foliation of $X$ is the group $G$ itself.


\section{The Laplacian on the deformed conifold}
\label{LAPLACIAN}

Before we embark on the discussion of the Laplace operator on the Stenzel space \eqref{DeformedConifold} with arbitrary $d$, we find it instructive to solve a simpler problem, namely $d = 4$.  There are a few advantages to studying this case first.   A relatively simple parametrization of this space in terms of angles is known \cite{Candelas:1989js}, and we will use it to obtain an expression for the Laplacian in terms of the $y_i$ coordinates; in section~\ref{GENERALIZATION} we will generalize this expression to arbitrary $d$.  In addition, the product structure of the isometry group $\grSO(4) \cong \grSU(2)_L \times \grSU(2)_R$ makes it easier to understand the group theory that stands behind the decoupling of the various harmonics.

\subsection{Coordinates and Killing vectors}
\label{T11PARAMETERIZATION}

For $V_{4, 2} = T^{1, 1}$, a standard parameterization of the $y_i$ coordinates is \cite{Candelas:1989js}
 \es{yForT11}{
  y_1 &= {1 \over \sqrt{2}} (a_1 b_1 - a_2 b_2)\; , \qquad  y_3 = -{1 \over \sqrt{2}} (a_1 b_2 + a_2 b_1)\; , \\
  y_2 &= {i \over \sqrt{2}} (a_1 b_1 + a_2 b_2)\; , \qquad y_4 = {i \over \sqrt{2}} (a_1 b_2 - a_2 b_1) \; ,
 }
where
 \es{abDefs}{
  a_1 &= \cos {\theta_1 \over 2} e^{i \left({\phi_1 \over 2} + {\psi \over 4} \right)}\; ,
   \qquad a_2 = \sin {\theta_1 \over 2} e^{i \left(-{\phi_1 \over 2}  + {\psi \over 4} \right)}\; , \\
  b_1 &= \cos {\theta_2 \over 2} e^{i \left({\phi_2 \over 2} + {\psi \over 4} \right)}\; ,
   \qquad b_2 = \sin {\theta_2 \over 2} e^{i \left(-{\phi_2 \over 2}  + {\psi \over 4} \right)} \; .
 }
The function ${\cal F}'(\tau)$ appearing in the metric \eqref{MetricAgain} of the deformed conifold is given by
 \es{FpT11}{
  {\cal F}'(\tau)
     = {\epsilon^{4/3}}(\cosh\tau\sinh\tau - \tau)^{1/3} \; .
 }

To describe the six $\grSO(4) \cong \grSU(2)_L \times \grSU(2)_R$ Killing vectors, it is convenient to split the generators $T^{(a)}$ into two groups corresponding to the left and the right $\grSU(2)$ factors.  The generators of $\grSU(2)_L$ are
 \es{SU2Left}{
  T^{(1)} &= {1 \over 2} \begin{pmatrix}
    0 & 1 & 0 & 0 \\
    -1 & 0 & 0 & 0 \\
    0 & 0 & 0 & 1 \\
    0 & 0 & -1 & 0
    \end{pmatrix}
   \qquad
  T^{(2)} = {1 \over 2} \begin{pmatrix}
    0 & 0 & 1 & 0 \\
    0 & 0 & 0 & -1 \\
    -1 & 0 & 0 & 0 \\
    0 & 1 & 0 & 0
    \end{pmatrix}
    \qquad
  T^{(3)} = {1 \over 2} \begin{pmatrix}
    0 & 0 & 0 & -1 \\
    0 & 0 & -1 & 0 \\
    0 & 1 & 0 & 0 \\
    1 & 0 & 0 & 0
    \end{pmatrix} \; ,
 }
and the generators of $\grSU(2)_R$ are
 \es{SU2Right}{
  T^{(4)} &= {1 \over 2} \begin{pmatrix}
    0 & 1 & 0 & 0 \\
    -1 & 0 & 0 & 0 \\
    0 & 0 & 0 & -1 \\
    0 & 0 & 1 & 0
    \end{pmatrix}
  \qquad
  T^{(5)} = {1 \over 2} \begin{pmatrix}
    0 & 0 & 1 & 0 \\
    0 & 0 & 0 & 1 \\
    -1 & 0 & 0 & 0 \\
    0 & -1 & 0 & 0
    \end{pmatrix}
   \qquad
  T^{(6)} = {1 \over 2} \begin{pmatrix}
    0 & 0 & 0 & 1 \\
    0 & 0 & -1 & 0 \\
    0 & 1 & 0 & 0 \\
    -1 & 0 & 0 & 0
    \end{pmatrix} \; .
 }
These generators
have been normalized so that they satisfy the algebra $[T^{(i)}, T^{(j)}] = \epsilon^{ijk} T^{(k)}$, $[T^{(i+3)}, T^{(j+3)}] = \epsilon^{ijk} T^{(k+3)}$, all other commutators being zero.  The Killing vectors can be computed straightforwardly from \eqref{KillingVectors}.  In terms of the angular coordinates, they are
 \es{KillingDeformedT11}{
  \xi^{(1)} &= - \frac\partial{\partial\phi_2} \; , \\
  \xi^{(2)} &= -\cos \phi_2 \frac\partial{\partial\theta_2}
   + \cot \theta_2 \sin \phi_2 \frac\partial{\partial\phi_2}
   - \csc \theta_2 \sin \phi_2 \frac\partial{\partial\psi} \; , \\
  \xi^{(3)} &= - \sin \phi_2 \frac\partial{\partial\theta_2}
   - \cot \theta_2 \cos \phi_2 \frac\partial{\partial\phi_2}
   + \csc \theta_2 \cos \phi_2 \frac\partial{\partial\psi} \; , \\
  \xi^{(4)} &= - \frac\partial{\partial\phi_1} \; , \\
  \xi^{(5)} &= -\cos \phi_1 \frac\partial {\partial \theta_1}
   + \cot \theta_1 \sin \phi_1 \frac \partial {\partial \phi_1}
   - \csc \theta_1 \sin \phi_1 \frac \partial{\partial \psi} \; , \\
  \xi^{(6)} &= - \sin \phi_1 \frac \partial{\partial \theta_1}
   - \cot \theta_1 \cos \phi_1 \frac\partial{\partial\phi_1}
   + \csc \theta_1 \cos \phi_1 \frac\partial{\partial\psi} \; .
 }
Quite nicely, the first three Killing vectors act by $\grSU(2)$ rotations of $b_1$ and $b_2$, while the last three act by $\grSU(2)$ rotations of $a_1$ and $a_2$.  We define the quadratic Casimir of $\grSO(4)$ to be
 \es{QuadCasimirDef}{
  {\cal C} = -2 \sum_{a = 1}^6 \xi^{(a)} \xi^{(a)} \; .
 }
Denoting by
 \es{JLRDefs}{
  J_{Li} = i \xi^{(i)}\; , \qquad J_{Ri} = i \xi^{(i+3)}\; , \qquad i = 1, 2, 3
 }
the angular momentum operators corresponding to $\grSU(2)_L$ and $\grSU(2)_R$, respectively, one can see that ${\cal C}$ can be written as
 \es{QuadCasimir}{
  {\cal C} = 2 (J_L^2 + J_R^2) \; .
 }
From the angular momentum algebra, it follows that functions that transform under the $[j_L,j_R]$ representation of $\grSO(4)$ have ${\cal C}$ eigenvalues
 \es{C2evalue}{
  {\cal C} = 2 \left[ j_L (j_L + 1) + j_R (j_R + 1) \right] \; .
 }
We will explain later how to construct such functions.   The reason for the factor of $2$ in \eqref{QuadCasimir} is that this way the usual $\grSO(4)$ spherical harmonics, for which $j_L = j_R = \ell / 2$, will have eigenvalues $\ell ( \ell + 2)$.

\subsection{The Laplacian}
\label{LAPLACIANCALCULATION}

The Laplacian on the deformed conifold can be decomposed into a sum of four terms according to their angular dependence:
 \es{LapDeformedT11}{
  \lap_6 = {\cal T} + g_{\cal C}(\tau) {\cal C} + g_{\cal R}(\tau) {\cal R} + g_{\cal L} (\tau) {\cal L} \; .
 }
The first term, \es{GotLrho}{
  {\cal T} &= \frac{6}{\epsilon^{4/3}\sinh^2\tau} \frac\partial{\partial\tau}\left((\cosh\tau\sinh\tau - \tau)^{2/3}\frac\partial{\partial\tau}\right) \; ,
 }
is a second order differential operator in the radial variable $\tau$. The others contain derivatives only with respect to the angular coordinates.  In terms of the complex coordinates $y_i$, they can be written as\footnote{
In writing down eq.~\eqref{GotLs}, we do not necessarily require that all coordinates $y_i$ and $\bar y_i$ should be treated as independent---one could solve, for example, for $y_1$, $\bar y_1$, and the real part of $y_2$ using \eqref{yConstraints} before applying any of the differential operators in \eqref{GotLs}.  But one doesn't have to.  Given any two functions $f_1(y_i, \bar y_i)$ and $f_2(y_i, \bar y_i)$ that are equal on the constrained surface \eqref{yConstraints}, the formulae in \eqref{GotLs} are written such that ${\cal D} f_1 = {\cal D} f_2$ on this constrained surface, where ${\cal D}$ can be either of the operators ${\cal C}$, ${\cal R}$, or ${\cal L}$.  That ${\cal D} f_1 = {\cal D} f_2$ can be checked by writing ${\cal D}$ explicitly in terms of angles using the formulae in section~\ref{T11PARAMETERIZATION}, and then applying the chain rule to convert all derivatives with respect to angles into derivatives with respect to $y_i$ and $\bar y_i$.}
\es{GotLs}{
  {\cal C} &= y_i y_j \frac{\partial^2}{\partial y_i\partial y_j}
    + (\bar y_i y_j - \delta_{ij} y_k \bar y_k)  \frac{\partial^2}{\partial y_i \partial \bar y_j}
    + 3 y_k \frac\partial{\partial y_k} + \text{c.c.} \\
   {\cal R} &= \left( y_i \frac\partial{\partial y_i}  - \bar y_i \frac\partial{\partial \bar y_i} \right)
   \left( y_j \frac\partial{\partial y_j}  - \bar y_j \frac\partial{\partial \bar y_j} \right) \\
  {\cal L} &=   {1\over 2} (\bar y_i y_j + y_i \bar y_j - \delta_{ij} y_k \bar y_k) \frac{\partial^2}{\partial y_i \partial y_j}
    + \bar y_k \frac\partial{\partial y_k} + \text{c.c.}
 }
The coefficients of these terms are functions of $\tau$:
 \es{Gotfs}{
  g_{\cal C}(\tau) = -{2 \coth\tau  \over  {\cal F}'} \; , 
  \qquad
  g_{\cal R}(\tau) = -{1 \over  {\cal F}''} + {2 \coth\tau \over {\cal F}'} \; ,
  \qquad
  g_{\cal L}(\tau) = {4 \csch\tau \over   {\cal F}'}
  \; ,
 }
with ${\cal F}'$ given in \eqref{FpT11}.  The operator ${\cal C}$ is the same as in \eqref{QuadCasimirDef}.  It can be checked that ${\cal C}$, ${\cal R}$, and ${\cal L}$ each commute with $\vec{J}_{L}$ and $\vec{J}_{R}$, so ${\cal C}$, ${\cal R}$, and ${\cal L}$ do not mix wavefunctions with different $\grSO(4)$ quantum numbers.

Our results agree with the formula for the Laplacian on the deformed conifold derived in \cite{Krishnan:2008gx} up to a typo: we believe that the function $A^2(\tau)$ = $2^{-10/3}\coth\frac\tau 2(\sinh 2\tau - 2\tau)^{1/3}$ defined in \cite{Krishnan:2008gx}
 should be $2^{-7/3}\coth\tau (\sinh 2\tau - 2\tau)^{1/3}$. The correspondence between our differential operators ${\cal C}$, ${\cal R}$, ${\cal L}$, ${\cal T}$ in equation \eqref{GotLs} and the operators $\square_R$, $\square_i$, $\square_m$, and $\square_\tau$ in \cite{Krishnan:2008gx} is
 \be
  {\cal C} = - 2 ( \square_1 +  \square_2 ) - 4 \square_R \; , \quad
  {\cal R} = -4 \square_R\, , \qquad {\cal L} = \square_m \; , \quad
  {\cal T} = \square_\tau \; .
 \ee
 We can also rewrite the coefficients $g_{\cal C}$, $g_{\cal R}$, and $g_{\cal L}$ from \eqref{Gotfs} as
 \be
  g_{\cal C} = -{1 \over \epsilon^{4/3}} {\coth^2 \tau \over 2 A(\tau)^2} \; , \quad
  g_{\cal R} = {1\over \epsilon^{4/3}}
     \left[ {\coth^2 \tau \over 2 A(\tau)^2} - {1 \over 4 B(\tau)^2} \right] \; ,
   \quad
  g_{\cal L} = {1 \over \epsilon^{4/3}} {\cosh \tau \over A(\tau)^2 \sinh^2 \tau} \;,
 \ee
where $B(\tau) = 2^{-1/3} 3^{-1} \sinh^2 \tau / (\sinh 2 \tau - 2 \tau)^{2/3} $, as in \cite{Krishnan:2008gx}.

\section{Green's Function}
\label{GREEN}

The results of the previous section enable us to find an expression for the Green's function of the Laplacian on the deformed conifold.
This question is of interest in gauge/string duality because it is relevant to the description of a stack of $p$ D3-branes on the deformed conifold, which describes one of the vacua of the $\grSU(p+ (k+1)M)\times \grSU(p+kM)$ gauge theory \cite{Klebanov:2000hb,Dymarsky:2005xt,Krishnan:2008gx}. In this case the 10-d metric is again of the form \eqref{typeIIBMetric}, but the warp factor is $H= H_{\rm KS} +\delta H$, where $H_{\rm KS}$ is the warp factor of the warped deformed conifold solution \cite{Klebanov:2000hb} and $\delta H$ is proportional to the Green's function with a source at the location $(\tau_0, \vec{y}_0)$ of the D3-branes \cite{Krishnan:2008gx}.  In particular, if $G$ is the Green's function normalized so that $\lap_6 G$ integrates to $1$ over the deformed conifold, $\delta H$ is given by \cite{Klebanov:2007us}
 \es{deltaHfromG}{
  \delta H(\tau, \vec{y}) = -(2\pi)^4 g_s p (\alpha')^2 G(\tau, \vec{y}; \tau_0, \vec{y}_0) \,,
 }
where $g_s$ is the string coupling constant and $1/(2\pi \alpha')$ is the string tension.  For simplicity, let us consider the D3-branes placed at a point on the blown up $S^3$ at $\tau=0$.\footnote{This problem was considered in \cite{Krishnan:2008gx} but our results are different; we find it necessary to include the mixing between different wave functions with the same
$\grSO(4)$ quantum numbers.}

\subsection{Normalization of the Green's function}

To normalize our Green's function, let us examine the form of the metric close to $\tau =0$.  With the complex coordinates $y_i$ written explicitly in terms of their real and imaginary parts as in \eqref{yDef}, the metric  at small $\tau$ takes the approximate form 
 \es{ApproxMetric}{
  ds_6^2 \approx {2^{1 \over 3} \epsilon^{4 \over 3} \over 3^{1\over 3}}
    \left[{1 \over 4} d\tau^2 + d\vec{u}\,^2 + {\tau^2 \over 4} d\vec{v}\,^2 \right]
    = {2^{1 \over 3} \epsilon^{4 \over 3} \over 3^{1\over 3}}
    \left[ d\vec{u}\,^2 + d\vec{w}\,^2 \right]\; ,
 }
where to obtain the last equality we introduced the coordinate $\vec{w} = \tau \vec{v}\, / 2$.  Geometrically, the vector $\vec{u}$ parameterizes the blown up $S^3$ at $\tau = 0$, and for a given $\vec{u}$, the vector $\vec{w}$ parameterizes the tangent $\R^3$ plane at $\vec{u}$, with $\abs{\vec{w}} = \tau/2$ being the radial coordinate in this tangent plane.

The equation satisfied by the Green's function of unit-strength source at $\vec{u} = \vec{u}_0$ and $\tau = 0$ is
 \es{Geq}{
  \lap_6 G(\tau, y_i; \vec{u}_0) = {3 \over 2 \epsilon^4}
    \delta^{(3)} (\vec{u} - \vec{u}_0)
    \delta^{(3)} (\vec{w})\; .
 }
The right-hand side of this equation is normalized so that it integrates to $1$ over the whole deformed conifold.  Note that both $\vec{u}$ and $\vec{w}$ were defined as vectors in $\R^4$, but each of them is forced to live on a three-dimensional surface because of the constraints $\abs{\vec{u}}^2 = 1$ and $\vec{w} \cdot \vec{u} = 0$.  The first delta-function in \eqref{Geq} is understood to integrate to $1$ over the unit $S^3$ parameterized by $\vec{u}$ with the standard metric on it;  the second delta-function is understood to integrate to $1$ over the $\R^3$ parameterized by $\vec{w}$, again assuming a standard metric.

\subsection{A coupled system of equations}

The solution to the Green's function problem \eqref{Geq} can be expanded in terms of polynomials in $y_i$ and $\bar y_i$ as in \eqref{Polyn} that have definite transformation properties under $\grSO(4)$.  The fact that the delta-function source is at $\tau = 0$ as opposed to a generic value of $\tau$ restricts drastically the number of $\grSO(4)$ irreps that contribute to this expansion.  To see which $\grSO(4)$ irreps contribute, one can use the symmetry of the problem as follows.

The group $\grSO(4)$ acts transitively on constant $\tau$ surfaces of the deformed conifold, and under the action of $\grSO(4)$ any point can be brought into the form
 \es{GenericPoint}{
   z_i = \epsilon \begin{pmatrix} 0 & 0 & i \sinh \frac{\tau}{2} & \cosh \frac{\tau}{2} \end{pmatrix}
 }
for some $\tau$.  A delta-function source at a generic point (for which $\tau>0$) therefore preserves an $\grSO(2)$ symmetry corresponding to rotations in the first two coordinates in \eqref{GenericPoint}.  This $\grSO(2)$ symmetry is enhanced to $\grSO(3)$ at $\tau = 0$, the $\grSO(3)$ acting on the first three coordinates in \eqref{GenericPoint}.  The Green's function corresponding to a delta-function source at a point on the $S^3$ at $\tau = 0$ can thus be expanded in harmonics invariant under an $\grSO(3)$ subgroup of $\grSO(4)$.   The embedding of this $\grSO(3)$ into $\grSO(4)$ depends on the precise location $\vec{u}_0$ of the delta-function source, and is generated by the linear combinations of the $\grSO(4)$ generators \eqref{JLRDefs} that, acting as Killing vectors, vanish at $\vec{u} = \vec{u}_0$.  A particularly simple choice corresponds to $\vec{u}_0 =(0,  0,  0,  1)$, where \eqref{KillingVectors} implies that the $\grSO(3)$ subgroup that leaves this point invariant is generated by $\vec{J}_L + \vec{J}_R$.\footnote{For a source at an arbitrary $\vec{u} = \vec{u}_0$ on $S^3$, the unbroken $\grSO(3)$ is generated by $O (\vec{J}_L + \vec{J}_R) O^T$, where $O$ is an $\grSO(4)$ matrix that rotates $(0,  0,  0,  1)$ into $( u_0^1,  u_0^2,  u_0^3,  u_0^4)$.  The matrix $O$ can be taken to be $u_0^4 I_4 + u_0^3 T^{(1)} - u_0^2 T^{(2)} - u_0^1 T^{(3)}$, where $I_4$ is the $4 \times 4$ identity matrix and the $T^{(i)}$ matrices were defined in \eqref{SU2Left}.}  With this choice, writing $\vec{J}_L$ and $\vec{J}_R$ as differential operators as in \eqref{JLRDefs}, one can see that only functions of $y_4$ and $\bar y_4$ are invariant under $\vec{J}_L + \vec{J}_R$.\footnote{Functions of $y_1^2 + y_2^2 + y_3^2$, $y_1 \bar y_1 + y_2 \bar y_2 + y_3 \bar y_3$, and $\bar y_1^2 + \bar y_2^2 + \bar y_3^2$ are also invariant, but they can be expressed as functions of $y_4$ and $\bar y_4$ using the constraints \eqref{yConstraints}.}

For the spin $[j_L, j_R]$ representation of $\grSU(2)_L \times \grSU(2)_R$, one can ask how many states are invariant under $\vec{J}_L + \vec{J}_R$.  Clearly, such a state should be invariant under $J_{L3} + J_{R3}$ so it should be a linear combination of $\ket{j_L, m_L}\ket{j_R,  -m_L}$.  Invariance under $J_{L\pm} + J_{R\pm}$ fixes $j_L = j_R$ and all coefficients in the linear combination to be proportional to $(-1)^{m_L}$.  So if $j_L \neq j_R$ there are no states invariant under $\vec{J}_L + \vec{J}_R$, and when $j_L = j_R$, only the state $\sum_{m_L = -j_L}^{j_L} (-1)^{m_L} \ket{j_L, m_L}\ket{j_L,-m_L}$ is invariant.  Quite nicely, this result can be reproduced from thinking about polynomials in $y_i$ and $\bar y_i$.  As already discussed, invariance under $\vec{J}_L + \vec{J}_R$ restricts these polynomials to be functions of $y_4$ and $\bar y_4$ only.  It is a straightforward exercise to check that $J_L^2 = J_R^2$ when acting on functions $F(y_4, \bar y_4)$, so all such functions are states in some $[j_L, j_R]$ irrep with $j_L = j_R$.  These states are not eigenstates of $J_{L3}$ or $J_{R3}$.

In section~\ref{FUNCTIONS} we explained why for a given spin $[j_L, j_R]$ irrep of $\grSO(4)$ there are precisely $2 \min (j_L, j_R) + 1$ functions on $T^{1, 1}$ with the same $\grSO(4)$ quantum numbers.  It follows that for each $\grSO(4)$ irrep with $j_L = j_R = \tilde \jmath$, there should in fact be $(2\tilde{\jmath}+1)$ linearly independent polynomials $F(y_4, \bar y_4)$ transforming in this irrep that are invariant under $\vec{J}_L + \vec{J}_R$.  One can find them by making the ansatz $F(y_4, \bar y_4) = y_4^{2 \tilde m} f( y_4 \bar y_4)$ or $F(y_4, \bar y_4) = \bar y_4^{2 \tilde m} f( y_4 \bar y_4)$, where $\tilde m$ was introduced at the end of section~\ref{FUNCTIONS}.  The equation $J_L^2 F(y_4, \bar y_4) = \tilde{\jmath}(\tilde{\jmath}+1) F(y_4, \bar y_4)$ then reduces to a hypergeometric differential equation for $f$.  Requiring regularity at all points on $T^{1, 1}$, one finds that the $(2\tilde{\jmath}+1)$ linearly independent polynomials are\footnote{These functions can be expanded in the basis of states described in the following section as \[F_{\tilde{\jmath}\tilde{m}} = (-1)^{\tilde{\jmath}} \frac{\sqrt{2\tilde{\jmath}+1}}{2^{\tilde{\jmath}}} \sqrt{\frac{(2\tilde{\jmath})!}{(\tilde{\jmath}-\tilde{m})!(\tilde{\jmath}+\tilde{m})!}} \sum_{m=-\tilde{\jmath}}^{\tilde{\jmath}} (-1)^{m} \ket{2\tilde{\jmath}; \tilde{\jmath} , m ; \tilde{\jmath} , -m ; \tilde{m}} \; .\]}
 \es{InvarPolyn}{
  F_{\tilde{\jmath} \tilde m}(y_4, \bar y_4) = {\cal N}_{\tilde{\jmath} \tilde m} \; {}_2 F_1 (-\tilde{\jmath} + \abs{\tilde m}, 1 + \tilde{\jmath} + \abs{\tilde m} ; 1 + 2 \abs{\tilde m}; 2 y_4 \bar y_4)
    \times \begin{cases}
   y_4^{2 \tilde m} & \text{if $\tilde m \geq 0$} \\
   \bar y_4^{2 \abs{\tilde m}} & \text{if $\tilde m < 0$}\; ,
   \end{cases}
 }
where $\tilde m$ takes values $-\tilde{\jmath}, -\tilde{\jmath}+1, \ldots, \tilde{\jmath}$, and ${\cal N}_{\tilde{\jmath} \tilde m}$ is a normalization constant that we take to be
 \es{GotNm}{
  {\cal N}_{\tilde{\jmath} \tilde{m}} = (-1)^{-\tilde{\jmath}} \, (-2)^{\abs{\tilde{m}}} \, \sqrt{2\tilde{\jmath}+1} \, \binom{\tilde{\jmath}+\abs{\tilde{m}}}{\tilde{\jmath}-\abs{\tilde{m}}} \; .
 }
This normalization is chosen such that
\es{FNorm}{
  \frac{1}{\Vol(T^{1,1})} \int_{T^{1,1}}\!d^5y \: F_{\tilde{\jmath}\tilde{m}}^* F_{\tilde{\jmath}'\tilde{m}'} = \delta_{\tilde{\jmath}\tilde{\jmath}'} \delta_{\tilde{m}\tilde{m}'} \; .
}
This formula is independent of the metric one chooses on $T^{1, 1}$ as long as the corresponding volume form is $\grSO(4)$-invariant.  The hypergeometric function appearing in \eqref{InvarPolyn} is just a polynomial in $y_4 \bar y_4$ of degree $\tilde{\jmath} - \abs{\tilde m}$; for example:
\be \label{symmmodes}
                 F_{0,0} = 1 \quad
\begin{array}{l} F_{\half,\half}  = 2 y_4  \\
                 F_{\half,-\half} = 2 \bar{y}_4 \end{array}  \quad
\begin{array}{l} F_{1,1}  = 2 \sqrt{3} \, y_4^2 \\
                 F_{1,0}  = - \sqrt{3} \, (1 - 4 y_4 \bar{y}_4) \\
                 F_{1,-1} = 2 \sqrt{3} \, \bar{y}_4^2 \end{array}  \quad
\begin{array}{l} F_{\frac{3}{2},\frac{3}{2}}  =  4 \sqrt{2} \, y_4^3 \\
                 F_{\frac{3}{2},\frac{1}{2}}  = -4 \sqrt{2} \, y_4 (1 - 3 y_4 \bar{y}_4) \\
                 F_{\frac{3}{2},-\frac{1}{2}} = -4 \sqrt{2} \, \bar{y}_4 (1 - 3 y_4 \bar{y}_4) \\
                 F_{\frac{3}{2},-\frac{3}{2}} =  4 \sqrt{2} \, \bar{y}_4^3 \,. \end{array}
\ee
With the normalization in \eqref{GotNm}, we can find how the differential operators \eqref{GotLs} act on the polynomials $F_{\tilde \jmath \tilde m}$:
 \es{OpF}{
  {\cal C} F_{\tilde{\jmath} \tilde m} &= 4 \tilde{\jmath}(\tilde{\jmath}+1) F_{\tilde{\jmath} \tilde m} \; ,
  \\
  {\cal R} F_{\tilde{\jmath} \tilde m} &= 4 \tilde m^2 F_{\tilde{\jmath} \tilde m} \; ,
  \\
  {\cal L} F_{\tilde{\jmath} \tilde m} &=
    (\tilde{\jmath} + \tilde m)(\tilde{\jmath} - \tilde m + 1) F_{\tilde{\jmath}, \tilde m-1}
   + (\tilde{\jmath} - \tilde m)(\tilde{\jmath} + \tilde m + 1) F_{\tilde{\jmath}, \tilde m+1} \; .
 }

The Green's function $G(\tau, y_i; \vec{u}_0)$ can be expanded in the basis of $F_{\tilde{\jmath} \tilde m}$ functions,
 \es{GreensFunction}{
  G(\tau, y_i; \vec{u}_0) = \sum_{\tilde{\jmath} = 0, {1 \over 2}, 1, \ldots} \sum_{\tilde m = -\tilde{\jmath}}^{\tilde{\jmath}} f_{\tilde{\jmath}\tilde m}(\tau) F_{\tilde{\jmath} \tilde m}(y_4, \bar y_4) \; ,
 }
with the coefficients $f_{\tilde{\jmath}\tilde{m}}(\tau)$ being functions of $\tau$.  Because the source in \eqref{Geq} preserves the $\Z_2$ symmetry of the deformed conifold, the $f_{\tilde{\jmath}\tilde{m}}(\tau)$ are not all independent. The $\Z_2$ symmetry acts on the coordinates by exchanging $y_i \leftrightarrow \bar{y}_i$, and therefore maps $F_{\tilde{\jmath} \tilde m}(y_4, \bar y_4) \to F_{\tilde{\jmath} \tilde m}(\bar y_4,y_4) = F_{\tilde{\jmath}, -\tilde m}(y_4, \bar y_4)$. Thus, for given $\tilde{\jmath}$ there are $n_{\tilde{\jmath}} = \bigceiling{\frac{2\tilde{\jmath}+1}{2}} = \bigfloor{\tilde{\jmath}}+1$ independent functions $f_{\tilde{\jmath}\tilde{m}}(\tau) = f_{\tilde{\jmath},-\tilde{m}}(\tau)$ with $\tilde{m}\ge0$.

Using \eqref{LapDeformedT11}, \eqref{FNorm}, and \eqref{OpF}, one can derive a system of coupled ordinary differential equations for these functions:
\be \label{ODEfjm}
  && {\cal T} f_{\tilde{\jmath}\tilde{m}}
     + \bigsbrk{ 4\tilde{\jmath}(\tilde{\jmath}+1) g_{\cal C} + 4 \tilde{m}^2 g_{\cal R} } f_{\tilde{\jmath}\tilde{m}} \nln[1mm]
  && \hspace{15mm} + (\tilde{\jmath}+\tilde{m}+1)(\tilde{\jmath}-\tilde{m}) g_{\cal L} f_{\tilde{\jmath},\tilde{m}+1} \nln
  && \hspace{15mm} + (\tilde{\jmath}-\tilde{m}+1)(\tilde{\jmath}+\tilde{m}) g_{\cal L} f_{\tilde{\jmath},\tilde{m}-1}
     = \sqrt{2\tilde{\jmath}+1} \, \frac{3}{2\pi^3\epsilon^4} \frac{\delta(\tau)}{\sinh^2\tau} \,,
\ee
where the differential operator ${\cal T}$ is given in \eqref{GotLrho} and $g_{\cal C}$, $g_{\cal R}$, and $g_{\cal L}$ are defined in \eqref{Gotfs}. The right hand side of \eqref{ODEfjm} follows from
\be
  \int_{T^{1, 1}}d^5 y\,  \frac{3}{2\epsilon^4} \delta^{(3)}(u-u_0)\,\delta^{(3)}(w) F^*_{\tilde{\jmath}\tilde{m}}(y_4,\bar{y}_4)
  \eq \frac{8}{9\epsilon^4} \frac{\delta(\tau)}{\sinh^2\tau} F^*_{\tilde{\jmath}\tilde{m}}(1/\sqrt{2},1/\sqrt{2})
\ee
and $F^*_{\tilde{\jmath}\tilde{m}}(1/\sqrt{2},1/\sqrt{2}) = \sqrt{2\tilde{\jmath}+1}$.  In evaluating the integral in the previous equation we have assumed that the measure on $T^{1, 1}$ is normalized so that the volume of $T^{1, 1}$ has the standard value $\Vol(T^{1,1}) = \frac{16}{27}\pi^3$ \cite{Herzog:2001xk}.  To see the coupling between different components $f_{\tilde{\jmath}\tilde{m}}$ explicitly, we write \eqref{ODEfjm} for $\tilde{\jmath} = 0, \half, 1$:
\be
  &&{\cal T} f_{0,0} = \frac{3}{2\pi^3\epsilon^4} \frac{\delta(\tau)}{\sinh^2\tau} \;, \\[2mm]
  &&{\cal T} f_{\half, \half} + (3g_{\cal C} + g_{\cal R} + g_{\cal L}) f_{\half, \half} = \frac{3\sqrt{2}}{2\pi^3\epsilon^4} \frac{\delta(\tau)}{\sinh^2\tau} \;, \\[2mm]
  &&  {\cal T} \matr{c}{ f_{1, 1} \\ f_{1, 0} }
  + \matr{cc}{ 8g_{\cal C} + 4g_{\cal R} & 2g_{\cal L} \\ 4g_{\cal L} & 8g_{\cal C} } \matr{c}{ f_{1, 1} \\ f_{1, 0} }  = \frac{3\sqrt{3}}{2\pi^3\epsilon^4} \frac{\delta(\tau)}{\sinh^2\tau} \;.
\ee

Let us now discuss the boundary conditions one has to impose in solving the equations for $f_{\tilde{\jmath}\tilde m}$.  Since for each $\tilde{\jmath}$ we are solving a system of $n_{\tilde{\jmath}} = \bigfloor{\tilde{\jmath}}+1$ second order differential equations, there are $2n_{\tilde{\jmath}}$ integration constants.  Near $\tau = 0$, eq.~\eqref{Geq} takes the form
 \es{GreenClose}{
    \left[\frac{1}{\tau^2} \frac{\partial}{\partial \tau} \tau^2 \frac{\partial}{\partial\tau}
    + \frac{-{\cal C} + {\cal R} + 2 {\cal L} }{2 \tau^2} \right] G(\tau, y_i; \vec{u}_0)
    = \frac{3^{2/3}}{2^{8/3} \epsilon^{8/3}} \delta^{(3)}(\vec{u} - \vec{u}_0) \delta^{(3)} (\vec{w}) \; ,
 }
where ${\cal C}$, ${\cal R}$, and ${\cal L}$ are $n_{\tilde{\jmath}} \times n_{\tilde{\jmath}}$ matrices.  Ignoring the delta-function source, eq.~\eqref{GreenClose} has power-law solutions that depend on the eigenvalues of the matrix $-{\cal C} + {\cal R} + 2 {\cal L}$.  It can be checked that these eigenvalues are $-4(k-2)(k-1)$ with $k = 1,\ldots,n_{\tilde{\jmath}}$, so there are $n_{\tilde{\jmath}}$ regular solutions that behave as $\tau^{2(k-1)}$ and $n_{\tilde{\jmath}}$ solutions that blow up as $1/\tau^{2k-1}$.  Because the first term in \eqref{GreenClose} is nothing but a $3$-d Laplacian in flat space and the RHS contains a $3$-d delta-function in $\vec{w}$, we infer that $G(\tau, y_i; \vec{u}_0) \sim 1/\tau$ at small $\tau$.  The precise normalization will be discussed shortly.  The requirement that $G(\tau, y_i; \vec{u}_0) \sim 1/\tau$ at small $\tau$ fixes $n_{\tilde{\jmath}}-1$ integration constants to zero, namely the coefficients of $1/\tau^{2k-1}$ with $k>1$.  In an asymptotic expansion of \eqref{Geq} as $\tau \to \infty$, there are again $2n_{\tilde{\jmath}}$ integration constants (of course, related linearly to the ones appearing in the $\tau \to 0$ expansion), half of them multiplying divergent terms and half of them multiplying solutions that vanish asymptotically.  Requiring regularity in this limit fixes an additional $n_{\tilde{\jmath}}$ integration constants, namely the coefficients of the divergent terms.  The last integration constant is fixed by imposing suitable normalization.  To sum up, for each $\tilde{\jmath}$ there is a unique solution to the $n_{\tilde{\jmath}}$ coupled ODEs in \eqref{Geq};  at small $\tau$ this solution behaves as $1/\tau$ with a coefficient that we will now determine.

The coefficients $c_{\tilde{\jmath}\tilde m}$ in the expansion
\es{CoeffsClose}{
  f_{\tilde{\jmath}\tilde{m}}(\tau) = \frac{c_{\tilde{\jmath}\tilde m}}{\tau} + \order(\tau^0)
}
must be designed so that the numerical factor and, in particular, the delta-functions on the right hand side of \eqref{GreenClose} are obtained when acting with the differential operator of \eqref{GreenClose} onto \eqref{GreensFunction}. The delta-function on $\R^3$ is readily produced by
\be
  \frac{1}{\tau^2} \frac{\partial}{\partial \tau} \tau^2 \frac{\partial}{\partial\tau} \, \frac{1}{\tau} = -\frac{\pi}{2} \delta^{(3)}(\vec{w}) \; .
\ee
Since the second term in \eqref{GreenClose} would introduce an unwanted $\tau$-dependence, we need to choose the coefficients such that the $1/\tau$ term of the Green's functions is annihilated by the operator $-{\cal C} + {\cal R} + 2 {\cal L}$. It can be checked that the kernel of this operator is spanned by the functions
\es{kernel}{
  E_{\tilde{\jmath}}(y_4,\bar{y}_4) =
  \sum_{\tilde{m}=-\tilde{\jmath}}^{\tilde{\jmath}} F_{\tilde{\jmath}\tilde{m}}(y_4,\bar{y}_4) \; .
}
Thus by choosing the coefficients $c_{\tilde{\jmath}\tilde m} = c_{\tilde{\jmath}}$ to be independent of $\tilde{m}$, we eliminate the second term and arrive at
 \es{GreenClose2}{
    \left[\frac{1}{\tau^2} \frac{\partial}{\partial \tau} \tau^2 \frac{\partial}{\partial\tau}
    + \frac{-{\cal C} + {\cal R} + 2 {\cal L} }{2 \tau^2} \right] G(\tau, y_i; \vec{u}_0)
    = -\frac{\pi}{2} \delta^{(3)} (\vec{w}) \sum_{\tilde{\jmath} = 0, {1 \over 2}, 1, \ldots} c_{\tilde{\jmath}} \, E_{\tilde{\jmath}}(y_4,\bar{y}_4) \; .
 }
In fact, the functions \eqref{kernel} depend only on the combination $u_4 = \frac{1}{\sqrt{2}} ( y_4 + \bar{y}_4 )$, as can be seen from the fact that the only normalizable functions annihilated by
 \es{Opuv}{
  -{\cal C} + {\cal R} + 2 {\cal L} =
  -4 v_4 {\partial \over \partial v_4} + 2 (1 - u_4^2 - v_4^2) {\partial^2 \over \partial v_4^2}
 }
are independent of $v_4 = \frac{1}{i\sqrt{2}} ( y_4 - \bar{y}_4 )$. If we denote $u_4 = \cos \theta$, the $E_{\tilde \jmath}$ can be written as
\es{kernal2}{
  E_{\tilde{\jmath}} = \sqrt{2 \tilde{\jmath}+1} \, \frac{\sin \lrsbrk{ (2 \tilde{\jmath}+1) \theta }}{\sin \theta} \; .
}
For our choice of $\vec{u}_0 = (0,0,0,1)$, the delta-function on $S^3$ can be decomposed as
\es{Completeness}{
  \delta^{(3)}(\vec{u} - \vec{u}_0) = \sum_{\tilde{\jmath} = 0, {1 \over 2}, 1, \ldots}^\infty
    \frac{\left(2 \tilde{\jmath}+1\right)}{2 \pi^2} \, \frac{\sin \lrsbrk{ (2 \tilde{\jmath}+1) \theta }}{\sin \theta} \; ,
}
so the right hand side of \eqref{GreenClose2} has just the required structure. Now, we can read off that the coefficients $c_{\tilde{\jmath}\tilde{m}}$ have to be
\es{coeffs}{
  c_{\tilde{\jmath}\tilde{m}} = c_{\tilde{\jmath}} = - \frac{\sqrt{2\tilde{\jmath}+1}}{\pi^3} \, \frac{3^{2/3}}{2^{8/3}\epsilon^{8/3}}
  }
in order to exactly reproduce \eqref{GreenClose}. This fully specifies the boundary conditions for the Green's function at small $\tau$.
Note that, upon summing over $\tilde{\jmath}$, the $1/\tau$ term cancels in the Green's function except at the location of the source; this is analogous to the cancelation of singularities in the Green's function on the resolved conifold \cite{Klebanov:2007us}.
Following the recipe described above, it is possible to solve for the functions $f_{\tilde{\jmath}\tilde{m}}(\tau)$ numerically. Such a numerical study is beyond the scope of the present paper.

\subsection{Expectation values of mesonic operators}

Placing $p$ mobile D3-branes on the warped deformed conifold provides the string dual of the cascading $\grSU((k+1)M+p)\times \grSU(kM +p)$ gauge theory \cite{Dymarsky:2005xt,Krishnan:2008gx}. The $3p$ complex coordinates of the D3-branes parameterize the mesonic moduli space of the gauge theory; they determine the expectation values of the mesonic operators. A special locus on the moduli space, which preserves an $\grSO(3)$ subgroup of the $\grSO(4)$ global symmetry, is to place all $p$ D3-branes at the same point of the $S^3$ at $\tau=0$. Their effect on the background is encoded in a shift of the warp factor proportional to the Green's function on the deformed conifold with a source at the location of the stack. The falloffs of various perturbations of the background at large $r\sim e^{\tau/3}$ determine the expectation values of various mesonic operators at this point on the moduli space.

From the coupled equations for the radial functions $f_{\tilde \jmath \tilde m}(\tau)$, we find that their leading behavior at large $\tau$ is
\es{leadingbeh}{f_{\tilde \jmath \tilde m} \sim C_{\tilde \jmath \tilde m} r^{-2\left[1+ \sqrt{1+3\tilde \jmath(\tilde \jmath+1)- 3{\tilde m}^2/4}\right]} \; ,
}
where $C_{\tilde \jmath\tilde m}$ are constants determined by numerical integration of the coupled ODEs\@. Since the unperturbed warp factor behaves as $(\ln r)r^{-4}$ at large $r$, we note that the perturbations of the transverse part of the metric, $r^2 H^{1/2}(r) d\omega_{T^{1,1}}^2$, fall off as
$r^{-\Delta_{\tilde \jmath\tilde m}} (\ln r)^{-1/2}$ where
\es{symdim}{
  \Delta_{\tilde \jmath\tilde m}=2\left[\sqrt{1+3\tilde \jmath(\tilde \jmath+1)- {{3\tilde m}^2 \over 4}}-1\right] \; .
}
Up to the power of $\ln r$ that appears due to the cascading behavior in the UV, this asymptotic behavior has the standard form corresponding to a VEV of an operator of dimension $\Delta_{\tilde \jmath\tilde m}$. Let us compare eq.~\eqref{symdim} with the formula for the dimensions of scalar operators dual to a mixture of the metric and 4-form perturbations with $\grSU(2)_L \times \grSU(2)_R \times \grU(1)_R$ quantum numbers $j_L, j_R, R$ \cite{Ceresole:1999zg}:
\es{alldim}{\Delta=2\left[\sqrt{1+{3 \over 2} j_L(j_L+1) + {3 \over 2} j_R (j_R+1) - {3 R^2 \over 16}}-1\right] \; .
}
The operators described by \eqref{symdim} have $j_L=j_R=\tilde \jmath$ and $\tilde m=R/2=-\tilde \jmath, -\tilde \jmath+1, \ldots, \tilde \jmath$. This is consistent with the fact that an $\grSO(3)$ global symmetry is preserved, but $\grU(1)_R$ is broken by the deformation of the conifold. Thus, only the $\grSO(3)$-invariant mesonic operators get VEVs. The form of these vertex operators in the gauge theory may be obtained from (\ref{symmmodes}) by replacing $y_4$ with
${i\over \sqrt 2}(A_1 B_2 - A_2 B_1)$ and taking an overall trace.
For $\tilde m=\tilde \jmath$ ($\tilde m=-\tilde \jmath$) these operators are chiral (anti-chiral) with protected dimensions $3|\tilde m|$, but for other values of $\tilde m$ the dimensions are not protected by supersymmetry.

\section{Basis functions on the deformed conifold}
\label{DECOUPLING}

In section \ref{FUNCTIONS}, we defined a basis of functions on $T^{1, 1}$ that consisted of linearly-independent polynomials $ M_{i_1 i_2 \cdots i_{n_1}}^{j_1 j_2 \cdots j_{n_2}} \, y_{i_1} y_{i_2} \cdots y_{i_{n_1}} \, \bar y_{j_1} \bar y_{j_2} \cdots \bar y_{j_{n_2}}$ defined up to the equivalence relations imposed by the constraints \eqref{yConstraints}.    In this section and the next we will find it convenient to represent each equivalence class by  a manifestly traceless polynomials constructed so that the tensor $M_{i_1 i_2 \cdots i_{n_1}}^{j_1 j_2 \cdots j_{n_2}}$ is traceless in any pair of the $n = n_1 + n_2$ indices.  Tracelessness in any two lower indices arises because the trace part is proportional to $\sum_{i = 1}^4 y_i^2 = 0$, and likewise for the upper indices.  The reason why we impose tracelessness in an upper and a lower index as well is that the trace part is proportional to $\sum_{i = 1}^4 \abs{y_i}^2 = 1$, giving a polynomial of lower degree.  As we will explain briefly, if we restrict ourselves to manifestly traceless polynomials we will be able to write down the operator ${\cal L}$ from section~\ref{LAPLACIANCALCULATION} in terms of the generators of an auxiliary $\widetilde{\grSU(2)}$ group.

\subsection{Group actions on traceless polynomials}

As we have shown in section~\ref{FUNCTIONS}, for the $[j_L, j_R]$ irrep of $\grSO(4) = \grSU(2)_L \times \grSU(2)_R$ there are precisely $2 \min(j_L, j_R) + 1$ linearly independent functions on $T^{1, 1}$ representing the same $\grSO(4)$ state.  Let us try to understand this last statement by examining in more detail the traceless polynomials in $y_i$ and $\bar y_i$.   First, one can check that the operator
 \es{CountDegree}{
  N = y_i \frac\partial{\partial y_i} + \bar y_i \frac\partial{\partial \bar y_i}
 }
which counts the degree of a monomial in $y_i$ and $\bar y_i$ commutes with any of the $J_{Li}$ and $J_{Ri}$ in the subspace of traceless polynomials. So it is consistent to restrict our attention to a subspace of homogeneous traceless polynomials of degree $n$.

A traceless tensor $M_{i_1 i_2 \cdots i_{n_1}}^{j_1 j_2 \cdots j_{n_2}}$ transforms in general in a \emph{reducible} representation of $\grSO(4)$. Its irreducible components differ by how many lower indices are symmetrized or anti-symmetrized with how many upper indices. We determine the occurring irreducible representations by building up $M$ from the product of $n_1$ fundamental representations ${\bf 4}_y$, representing the lower indices, and $n_2$ fundamental representations ${\bf 4}_{\bar y}$ for the upper indices. Writing $\grSO(4)$ as $\grSU(2)_L \times \grSU(2)_R$, one can write ${\bf 4}_y = [\half,\half]_y$, so the traceless symmetric product of $n_1$ ${\bf 4}_y$'s is $[\frac{n_1}{2},\frac{n_1}{2}]_y$. Similarly, we keep only the traceless symmetric product of $n_2$ ${\bf 4}_{\bar y}$'s, which gives the irreducible representation $[\frac{n_2}{2},\frac{n_2}{2}]_{\bar{y}}$. Lastly, we should only keep the traceless part of the product of these representations, which transforms as
 \es{FinalDecomp}{
\lrsbrk{\frac{n}{2},\frac{n}{2}} \oplus
\biggbrk{ \lrsbrk{\frac{n}{2},\frac{n}{2}-1} \oplus \lrsbrk{\frac{n}{2}-1,\frac{n}{2}} }
\oplus\cdots\oplus
\biggbrk{ \lrsbrk{\frac{n}{2},\frac{\abs{n_1-n_2}}{2}} \oplus \lrsbrk{\frac{\abs{n_1-n_2}}{2},\frac{n}{2}} } \; .
}

A decomposition like \eqref{FinalDecomp} appears for any partition of the degree $n$ of the polynomials \eqref{Polyn} into two integers, $n = n_1 + n_2$.  We therefore conclude that such traceless polynomials of degree $n$ are organized into $\grSO(4)$ irreps $[\frac{n}{2},\frac{n}{2}-k]$ and $[\frac{n}{2}-k,\frac{n}{2}]$, where $k$ is an integer satisfying $0 \leq k \leq \lrfloor{{n+1 \over 2}}$ and each representation appears precisely $n + 1 - 2k$ times. In particular, ${\cal C}$, ${\cal R}$, and ${\cal L}$ can mix together up to $n + 1 - 2k$ functions at a time.

The $n+1-2k$ distinct sets of polynomials differ only in the relative number of $y_i$'s and $\bar y_i$'s, and we can rotate between them with an $\widetilde{\grSU(2)}$ having the generators
 \es{SU2tGenerators}{
  \tilde J_1 &= {1 \over 2} \left[y_i \frac\partial{\partial \bar y_i} + \bar y_i \frac\partial{\partial y_i} \right]\; , \quad
  \tilde J_2 = {1 \over 2} \left[-i y_i \frac\partial{\partial \bar y_i} + i \bar y_i \frac\partial{\partial y_i} \right]\; , \quad
  \tilde J_3 = {1 \over 2} \left[y_i \frac\partial{\partial y_i} - \bar y_i \frac\partial{\partial \bar y_i} \right] \; .
 }
 Here $\tilde J_3$ counts (up to normalization) the number of $y_i$ minus the number of $\bar y_i$. Its eigenvalue $\tilde m$ running from $- {n \over 2} + k$ to ${n \over 2} - k$ is a magnetic quantum number that labels the distinct copies of each $\grSO(4)$ representation.  From $\tilde J_1$ and $\tilde J_2$ we can construct raising and lowering operators
 \es{Jtraisinglowering}{
  \tilde J_+ = y_i \frac\partial{\partial \bar y_i}\; , \qquad
   \tilde J_- = \bar y_i \frac\partial{\partial y_i} \; .
 }
Roughly speaking, $\tilde J_+$ turns one $\tilde y_i$ into $y_i$, while $\tilde J_-$ turns one $y_i$ into $\bar y_i$. Of course, the $\widetilde{\grSU(2)}$ action does not correspond to an isometry of the deformed conifold. It is nevertheless useful because, given a wavefunction of a particular $\tilde m$, one can use the operators \eqref{Jtraisinglowering} to construct those wavefunctions it potentially mixes with.

One can check that $N$, $J_L^2$, $J_{L3}$, $J_R^2$, $J_{R3}$, and $\tilde J_3$ form a maximally commuting set of operators in the subspace of traceless polynomials, so they can be simultaneously diagonalized.  The operator $\tilde J^2$ also commutes with all of these operators, but it is not independent because
 \es{GotJtSQ}{
  \tilde J^2 = J_L^2 + J_R^2 - {1 \over 4} N (N + 2) \; .
 }
 Homogeneous traceless polynomials $F_\alpha$ can therefore be described by six quantum numbers
 \es{AState}{
  \alpha = \ket{ n; j_L, m_L; j_R, m_R; \tilde m } \; ,
 }
where $n$ represents the degree of the polynomial, $j_L$ and $m_L$ are $\grSU(2)_L$ quantum numbers, $j_R$ and $m_R$ are $\grSU(2)_R$ quantum numbers, and $\tilde m$ is the eigenvalue of $\tilde J_3$, or the magnetic quantum number of the $\widetilde{\grSU(2)}$.  Given such a state, one can construct states with other values of $m_L$ and $m_R$ by acting with the raising and lowering operators $J_{L\pm} = J_{L1} \pm i J_{L2}$ and $J_{R\pm} = J_{R1} \pm i J_{R2}$.  One can also construct states with other values of $\tilde m$ by acting with the $\tilde J_+$ and $\tilde J_-$ operators defined in \eqref{Jtraisinglowering}.  Standard conventions read
 \es{RaiseLowerState}{
   \ket{ n; j_L, m_L; j_R, m_R; \tilde m \pm 1 } =
     { 1 \over \sqrt{\tilde{\jmath} (\tilde{\jmath} + 1)
     -\tilde m (\tilde m \pm 1)}} \tilde J_{\pm} \ket{ n; j_L, m_L; j_R, m_R; \tilde m } \; ,
 }
where as can be seen from \eqref{GotJtSQ},
 \es{Gotjt}{
  \tilde{\jmath} (\tilde{\jmath} + 1) = j_L (j_L + 1) + j_R (j_R + 1) - {n \over 2} \left({n \over 2} + 1\right) \; .
 }
For us, the discussion around \eqref{FinalDecomp} implies
 \es{jtSimp}{
  {n \over 2} = \max(j_L, j_R)
  \qquad \Longrightarrow \qquad \tilde \jmath = \min (j_L, j_R) \; .
 }

\subsection{Homogeneity properties of traceless polynomials}

To understand the significance of $\widetilde{\grSU(2)}$, it is useful to describe the homogeneity properties of the traceless polynomials on $T^{1, 1} = V_{4, 2}$ as discussed in refs.~\cite{Gelbart1, Gelbart2}.  Since we restricted ourselves to explicitly traceless polynomials, we can forget about the constraints \eqref{yConstraints}, as any equivalence class of polynomials with respect to the relations imposed by these constraints has an explicitly traceless representative.  So we can think of the $y_i$ and their complex conjugates $\bar y_j$ as unrestricted complex variables.  Actually, we can replace $\bar y_j$ by a new set of complex variables $\tilde y_j$ that are unrelated to $y_i$, and consider all manifestly traceless polynomials $P_{j_L, j_R} (y_i,\tilde y_j)$ in the $4 + 4 = 8$ independent complex variables $y_i$ and $\tilde y_j$.  When both $\tilde y_j = \bar y_j$ and the constraints \eqref{yConstraints} are satisfied, these polynomials would reduce to the functions on the Stiefel manifold $V_{4, 2}$.

The representation theory discussed above shows that $P_{j_L, j_R} (y_i, \tilde y_j)$ transforms in an irrep of $\grSO(4)$ whose Young diagram has two rows with $j_L + j_R$ boxes in the first row and $\abs{j_L - j_R}$ boxes in the second row.  The reason why only Young diagrams with two rows are possible is that any polynomial in $y_i$ or $\tilde y_j$ is automatically symmetric, so the only possible anti-symmetrizations are between $y_i$ and $\tilde y_j$, for example $y_1 \tilde y_2 - \tilde y_1 y_2$;  one simply cannot write down an anti-symmetric product of three or more $y_i$ and/or $\tilde y_j$.  Given such a Young diagram, it must be that each of the first $\abs{j_L - j_R}$ columns corresponds to either $y_i$ in the first row and $\tilde y_j$ in the second row, or the other way around.  Each of the remaining $2 \min(j_L, j_R)$ boxes in the first row corresponds to either a $y_i$ or a $\tilde y_j$, each choice giving a linearly-independent polynomial.  For this Young diagram there are therefore $2\min(j_L, j_R) + 1$ linearly-independent polynomials with the same $\grSO(4)$ quantum numbers, in agreement with the discussion from section~\ref{FUNCTIONS}.

We can now characterize the homogeneity properties of $P_{j_L, j_R} (y_i, \tilde y_j)$, following refs.~\cite{Gelbart1, Gelbart2}.  Thinking of $y_i$ and $\tilde y_j$ as the columns of a $4 \times 2$ matrix, we consider the $\grGL(2, \C)$ action
 \es{GL2CAction}{
   \begin{pmatrix} y_i & \tilde y_i \end{pmatrix} \to
   \begin{pmatrix} y_i'& \tilde y_i'\end{pmatrix} =
   \begin{pmatrix} y_i & \tilde y_i \end{pmatrix} M \; ,
 }
where $M$ is a $\grGL(2, \C)$ matrix.  For a given $\grSO(4)$ irrep, the $P_{j_L, j_R} (y_i, \tilde y_j)$ transform under the action \eqref{GL2CAction} as a representation $\rho_{j_L, j_R}$ of $\grGL(2, \C)$ because the various $P_{j_L, j_R} (y_i, \tilde y_j)$ have the same total number of $y_i$ and $\tilde y_j$.  Therefore
 \es{GL2CRep}{
    \begin{pmatrix} P_{j_L, j_R} (y_i',\tilde y_j') \end{pmatrix}
  = \begin{pmatrix} P_{j_L, j_R} (y_i, \tilde y_j ) \end{pmatrix}
  \rho_{j_L, j_R} (M) \; ,
 }
$\begin{pmatrix} P_{j_L, j_R} (y_i', \tilde y_j') \end{pmatrix}$ being a column vector representing the components of an arbitrary polynomial in the $[j_L, j_R]$ irrep of $\grSO(4)$ with respect to a basis of $2\min(j_L, j_R) + 1$ linearly-independent polynomials.  The group $\grGL(2, \C)$ has a maximally compact $\grU(2)$ subgroup, and $\rho_{j_L, j_R}$ can be restricted to a representation of $\grU(2)$ that we also denote by $\rho_{j_L, j_R}$.  One of the main results of refs.~\cite{Gelbart1, Gelbart2} is that $\rho_{j_L, j_R}$ is an irreducible $\grU(2)$ representation with the same Young diagram as the $[j_L, j_R]$ irrep of $\grSO(4)$, namely two rows, one with $j_L + j_R$ boxes and one with $\abs{j_L - j_R}$ boxes.  As a consistency check, it is easy to verify that the dimension of this irrep is $2 \min(j_L, j_R) + 1$, as expected.

From the action \eqref{GL2CAction} with $M$ restricted to be a unitary matrix, it follows that the generators of $\grU(2) \cong \widetilde{\grSU(2)} \times \grU(1)_N$ are $\tilde J_i$ for the $\widetilde{\grSU(2)}$ factor and $N$ for the $\grU(1)_N$ factor defined as in eqs.~\eqref{CountDegree} and~\eqref{SU2tGenerators}, but with the replacement $\bar y_i \to \tilde y_i$.  The $\grU(2)$ irrep described above restricts to a spin-$\tilde \jmath$ irrep of $\widetilde{\grSU(2)}$, where $\tilde \jmath = \min (j_L, j_R)$, and to the $\grU(1)_N$ irrep $e^{i \theta} \to e^{2 i \theta \max(j_L, j_R)}$, the latter restriction implying that all polynomials in the $[j_L, j_R]$ irrep have degree $n = 2 \max (j_L, j_R)$.  The values of $\tilde \jmath$ and $n$ that we just found from arguments about homogeneity properties agree with those found in eq.~\eqref{jtSimp} from a more brute-force method.

We should stress that $\grGL(2, \C)$ and its $\grU(2)$ subgroup do {\em not} correspond to symmetries of the Stenzel space---one can use the action of these groups on traceless polynomials just as a means of keeping track of the homogeneity  properties of these polynomials.  Indeed, except for the $\widetilde{\grU(1)}_3$ subgroup in $\widetilde{\grSU(2)}$ generated by $\tilde J_3$ and the $\Z_2$ subgroup that exchanges $y_i$ with $\tilde y_i$, the action of $\grU(2)$ does not even preserve the constraints \eqref{yConstraints}.  The action of $\widetilde{\grU(1)}_3$ is an R-symmetry when $\epsilon = 0$, but this symmetry is broken to $\Z_2$ for the deformed conifold.

\subsection{Explicit formulae for the Laplacian}

With the exception of ${\cal L}$ which fails to commute with $\tilde J_3$, the operators ${\cal C}$, ${\cal R}$, and ${\cal L}$ of \eqref{GotLs} all commute with $N$, $J_L^2$, $J_{L3}$, $J_R^2$, $J_{R3}$, and $\tilde J_3$.  We will therefore fix $n$, $j_L$, $m_L$, $j_R$, and $m_R$ and look at the $(2 \tilde{\jmath} + 1)$-dimensional subspace indexed by $\tilde m$.  Restricted to this subspace, ${\cal C}$ will have a fixed eigenvalue, ${\cal R}$ will be diagonal, and ${\cal L}$ will generically have off-diagonal entries.  In fact the following identities hold when acting on traceless tensors:
 \es{Identities}{
   {\cal C} = 2 (J_L^2 + J_R^2)\; , \qquad
   {\cal R} = 4 \tilde J_3^2\; , \qquad
   {\cal L} = - 2 i \tilde J_2 \tilde J_3 + \tilde J_1 N \; .
 }
The only non-vanishing matrix elements of these operators are
 \es{MatrixElements}{
  \langle n; j_L, m_L; j_R, m_R; \tilde m
    \,|\, {\cal C} \,|\,  n; j_L, m_L; j_R, m_R; \tilde m \rangle
     &= 2 \left[j_L (j_L + 1) + j_R (j_R + 1) \right] \; ,\\
   \langle n; j_L, m_L; j_R, m_R; \tilde m
    \,|\, {\cal R} \,|\,  n; j_L, m_L; j_R, m_R; \tilde m \rangle
     &=  4 \tilde m^2\; , \\
   \langle n; j_L, m_L; j_R, m_R; \tilde m \pm 1
    \,|\, {\cal L} \,|\,  n; j_L, m_L; j_R, m_R; \tilde m \rangle
     &=  \left({n \over 2} \mp \tilde m\right)
     \sqrt{(\tilde{\jmath} \mp \tilde{m}) (\tilde{\jmath} \pm \tilde{m} + 1)}
    \; ,
 }
where $\tilde \jmath = \min(j_L, j_R)$ as in \eqref{jtSimp}.

\subsection{Case studies}
\label{CASE}

The highest weight state in the $[j_L,j_R]$ representation of $\grSO(4)$, which is also a highest weight state of the spin $\tilde \jmath = \min(j_L, j_R)$ representation of $\widetilde{\grSU(2)}$, is
 \es{HighestWeight}{
  h_{j_L, j_R} = (y_2 + i y_3)^{2 \tilde \jmath}
   \left[(y_2 + i y_3)(\bar y_1 + i s \bar y_4) - (\bar y_2 + i \bar y_3) (y_1 + i s y_4) \right]^{\abs{j_L - j_R}}
 }
where $s = \sgn(j_L - j_R)$.  It corresponds to $F_\alpha$ with $\alpha = |2 \max(j_L, j_R) ; j_L , j_L ; j_R, j_R; \tilde \jmath\rangle$.  It is easy to check that $h_{j_L, j_R}$ is annihilated by $J_{L+}$ and $J_{R+}$ and that it is an eigenfunction of ${\cal C}$ with eigenvalue \eqref{C2evalue}.
Acting with $\tilde J_-$ several times on \eqref{HighestWeight} yields the highest-weight states in all the copies of the $\grSO(4)$ irrep $[j_L, j_R]$ indexed by $\tilde m$:
\be
  F_{j_L, j_R,\tilde{m}} \eq \sqrt{ { 2\tilde \jmath  \choose \tilde \jmath - \tilde{m} } }
                                \, (y_2 + i y_3)^{\tilde \jmath+\tilde{m}}
                                \, (\bar{y}_2 + i \bar{y}_3)^{\tilde \jmath -\tilde{m}} \nl
                                \hspace{13mm} \times \bigsbrk{ (y_2 + i y_3)(\bar{y}_1 + i s \bar{y}_4) - (y_1 + i s y_4)(\bar{y}_2 + i \bar{y}_3) }^{\abs{j_L - j_R}} \; .
\ee
Let us now give a few explicit examples.

\paragraph{Degree \mathversion{bold}$n = 1$.} In this case, there is only one possible $\grSO(4)$ representation  $[\half,\half]$.  The $\widetilde{\grSU(2)}$ spin is $\tilde{\jmath} = 1/2$, so there are two sets of states that mix with each other.  For example, one can consider the highest weight $\grSO(4)$ states $F_{{1\over 2}, {1\over 2}, \tilde m}$:
 \es{2times2}{
  F_{{1\over 2}, {1\over 2}, {1 \over 2}} = y_2 + i y_3 \; , \qquad
  F_{{1\over 2}, {1\over 2}, -{1\over 2}} = \bar y_2 + i \bar y_3 \; .
 }
In this two-dimensional subspace we have
 \es{nOneOps}{
  {\cal C} = 3 \; , \qquad {\cal R} = 1\; ,
   \qquad {\cal L} = \begin{pmatrix}
    0 & 1 \\
    1 & 0
   \end{pmatrix} \; .
 }

\paragraph{Degree \mathversion{bold}$n = 2$.} In this case, there are three possible $\grSO(4)$ representations: $[1,1]$, $[1,0]$, and $[0,1]$.  The representation $[1,1]$ has $\widetilde{\grSU(2)}$ spin $\tilde{\jmath} = 1$, so there are three function that mix.  The $\grSO(4)$ highest weight states $F_{1, 1, \tilde m}$ are
 \es{3times3}{
   F_{1, 1, 1} = (y_2 + i y_3)^2 \; , \qquad
   F_{1, 1, 0} = \sqrt{2} (y_2 + i y_3) (\bar y_2 + i \bar y_3) \; , \qquad
   F_{1, 1, -1} = (\bar y_2 + i \bar y_3)^2 \; .
 }
In this three-dimensional subspace,
 \es{nTwoOps1}{
  {\cal C} = 8 \; , \qquad
   {\cal R} = \diag \left\{4, 0, 4 \right\} \; ,
    \qquad
   {\cal L} = \begin{pmatrix}
    0 & \sqrt{2} & 0 \\
    2 \sqrt{2} & 0 & 2 \sqrt{2} \\
    0 & \sqrt{2} & 0
   \end{pmatrix} \; .
 }

The representations $[1,0]$ and $[0,1]$ are related by a $\Z_2$-symmetry, so we will only focus on $[1,0]$.  Here $\tilde{\jmath} = 0$, so there is no mixing.  The highest weight state is
 \es{3times1}{
  F_{1, 0, 0} = (y_2 + i y_3)(\bar y_1 + i \bar y_4) - (\bar y_2 + \bar y_3) (y_1 + i y_4) \; .
 }
Then
 \es{nTwoOps2}{
  {\cal C}= 4 \; , \qquad {\cal R} = 0 \; ,
   \qquad {\cal L} = 0 \; .
 }

\paragraph{Degree \mathversion{bold}$n = 3$.} The possible representations are $[\frac{3}{2},\frac{3}{2}]$, $[\frac{3}{2},\frac{1}{2}]$, and $[\frac{1}{2},\frac{3}{2}]$.  In the case $[\frac{3}{2},\frac{3}{2}]$, the $\grSO(4)$ highest weight states $F_{\frac 32, \frac 32, \tilde m}$ are
 \es{4times4}{
   F_{{3\over 2}, {3 \over 2}, {3 \over 2}} &= (y_2 + i y_3)^3 \; , \qquad
   F_{{3\over 2}, {3\over 2}, {1 \over 2}} = \sqrt{3} (y_2 + i y_3)^2 (\bar y_2 + i \bar y_3) \; ,
   \\
   F_{{3\over 2}, {3\over 2}, -{1 \over 2}} &= \sqrt{3} (y_2 + i y_3) (\bar y_2 + i \bar y_3)^2 \; , \qquad
   F_{{3\over 2}, {3\over 2}, -{3 \over 2}} = (\bar y_2 + i \bar y_3)^3 \; .
 }
The differential operators ${\cal C}$, ${\cal R}$, and ${\cal L}$ evaluate to
 \es{nThreeOps1}{
  {\cal C} = 15 \; , \qquad
   {\cal R} = \diag \left\{9, 1, 1, 9 \right\} \; ,
   \qquad
   {\cal L} = \begin{pmatrix}
    0 & \sqrt{3} & 0 & 0 \\
    3 \sqrt{3} & 0 & 4 & 0 \\
    0 & 4 & 0 & 3 \sqrt{3} \\
    0 & 0 & \sqrt{3} & 0
   \end{pmatrix} \; .
 }

For $[\frac{3}{2},\frac{1}{2}]$ the functions $F_{\frac 32, \frac 12, \tilde m}$ corresponding to the highest weight states are
 \es{4times2}{
  F_{{3\over 2}, {1 \over 2}, {1 \over 2}} &= (y_2 + i y_3) \left[(y_2 + i y_3) (\bar y_1 + i \bar y_4) - (\bar y_2 + i \bar y_3) (y_1 + i y_4) \right] \; , \\
  F_{{3\over 2}, {1\over 2}, -{1 \over 2}} &= (\bar y_2 + i \bar y_3) \left[(y_2 + i y_3) (\bar y_1 + i \bar y_4) - (\bar y_2 + i \bar y_3) (y_1 + i y_4) \right] \; .
 }
Then
 \es{nThreeOps2}{
  {\cal C} = 9\; , \qquad {\cal R} = 1\; ,
  \qquad
  {\cal L} = \begin{pmatrix}
   0 & 2 \\
   2 & 0
  \end{pmatrix} \; .
 }

%

\section{Glueball masses for the type IIB background}
\label{GLUEBALLSIIB}

\subsection{General procedure and case studies}

As was already discussed, the spectrum of spin-2 glueballs is given by those values of $m^2$ for which there exist normalizable solutions to the ten-dimensional Laplace equation $\lapten \Phi = 0$. Under the plane wave ansatz in the non-compact directions $x_\mu$, this equation reduces to the eigenvalue problem \eqref{EValue}
\es{EValue2}
 {\lap_6\phi = -m^2H(\tau)\phi \; ,
}
where
\es{htau}{
  H(\tau) = {2 (g_s M \alpha')^2 \over \epsilon^{8/3}} \int_\tau^\infty dx \frac{x\coth x -1}{\sinh^2 x}(\sinh x \cosh x - x)^{1/3} \; ,
}
and $\phi$ is a function of the internal coordinates $\tau$ and $(y_i, \bar y_j)$.  The component of $\phi$ along a state in the spin $[j_L,j_R]$ representation of $\grSO(4)$ with magnetic quantum numbers $m_L$ and $m_R$ has the form
\es{phif}{
  \phi =
    \sum_{\tilde m = -\tilde \jmath}^{\tilde \jmath} f_{\tilde m}(\tau)F_{\tilde m} (y_i, \bar y_i) \; .
}
The equations satisfied by the coefficient functions $f_{\tilde m}(\tau)$ are independent of the values of $m_L$ and $m_R$, but of course depend on $j_L$ and $j_R$.  In the previous section we derived a matrix representation of $\lap_6$ in the basis of the highest-weight polynomials $F_{j_L, j_R, \tilde{m}}$. So the problem is reduced to solving a system of coupled ODEs for the coefficient functions $f_{\tilde m}(\tau)$.

These equations decouple in the UV, where the eigenmodes are approximately those of $AdS_5 \times T^{1,1}$, up to the effects of the cascade. The latter spectrum was identified in earlier work \cite{Gubser:1998vd, Ceresole:1999zs} as follows: a wavefunction $\phi_i$, carrying quantum numbers $|n; j_L; m_L; j_R; m_R; \tilde m\rangle$, behaves asymptotically as
\es{asymptotics}{
  \phi_i \sim 
  c_{i} \left( \exp\left[ {(\Delta-4) \frac {\tau}{3}}\right] + \ldots \right)+ 
  d_{i} \exp\left[- \Delta \frac {\tau}{3} \right] \; ,
}
up to powers of $\tau$. The normalizable states have $c_i = 0$.  Then the asymptotics in eq.~\eqref{asymptotics} correspond to VEV's of operators of dimension \cite{Ceresole:1999zs}
\es{hdef}{
  \Delta =  2 + \sqrt{4 + 6[j_L(j_L+1) + j_R(j_R+1)] - 3\tilde m^2 } \; .
}
Note that there is no dependence on $n, m_L, m_R$.

The eigenmodes within each $\grSO(4)$ representation can be decomposed into an even and odd sector under the $\Z_2$ parity symmetry that interchanges $y_i$ with $\bar{y}_i$.  From \eqref{zToy}, it follows that interchanging $y_i$ and $\bar{y}_i$ is equivalent to sending $\tau \to -\tau$.  Consequently, the parity of the modes is manifested in the parity of the radial wavefunctions $f_{\tilde m}(\tau)$, where now $\tau$ is allowed to run from $-\infty$ to $\infty$. For the $[j_L,j_R]$ representation, the number of even modes is $n_e = \big\lceil\frac{2 \tilde \jmath + 1} 2 \big\rceil$ and the number of odd modes is $n_o = \big\lfloor \frac {2 \tilde \jmath + 1} 2 \big\rfloor$, where
$\tilde \jmath = \min(j_L, j_R)$.

Let us discuss some examples explicitly. For representations of the form $[j_L,0]$, we have $\tilde{\jmath} = 0$ and therefore no mixing. The wave function has only one component, $\phi(\tau,y_i,\bar{y}_i) = f_0(\tau) F_{j,0,0}(y_i,\bar{y}_i)$, and \eqref{EValue2} leads to
\be
  {\cal T} f_0 + \bigsbrk{ 2j_L(j_L+1) g_{\cal C} + m^2 H } f_0 = 0 \; .
\ee
For the representations $[\half,\half]$, the wave function has two components, $f_{\pm1/2}$, which are coupled by \eqref{EValue2}. In terms of $f^\pm_{1/2} = f_{1/2} \pm f_{-1/2}$ the equations decouple and we have
\be
  {\cal T} f^\pm_{1/2} + \bigbrk{ 3 g_{\cal C} + g_{\cal R} \pm g_{\cal L} + m^2 H } f^\pm_{1/2} = 0 \; .
\ee
The wave function for representation $[1,1]$ has the three components $f_1$, $f_0$, $f_{-1}$ according to $\tilde\jmath=1$. The first truly coupled equations appear in the even sector between the functions $f^+_1 = f_1 + f_{-1}$ and $f^+_0 = f_0$:
\be
  {\cal T} \matr{c}{f^+_1 \\ f^+_0} + \matr{cc}{ 8 g_{\cal C} + 4 g_{\cal R} + m^2 H & 2\sqrt{2} g_{\cal L} \\ 2\sqrt{2} g_{\cal L} & 8 g_{\cal C} + m^2 H } \matr{c}{f^+_1 \\ f^+_0} = 0 \; ,
\ee
while in the odd sector there is only one equation for $f^-_1 = f_1 - f_{-1}$:
\be
  {\cal T} f^-_1 + \bigbrk{ 8 g_{\cal C} + 4 g_{\cal R} + m^2 H } f^-_1 = 0 \; .
\ee
Because of the mixing of modes, in a typical glueball state several operators of different dimensions acquire expectation values. For example, for the parity even $j_L=j_R=1$ glueballs the large $\tau$ asymptotics, $f_1^+\sim e^{-7\tau/3}$ and $f_0^+\sim e^{-2(1+\sqrt 7)/3}$, show that both operators of dimension 7 and dimension $2(1+\sqrt 7)$ have expectation values.

\subsection{Numerical results}

To solve the coupled equations, it is convenient to use the ``determinant method" of \cite{Berg:2006xy}.  Let us focus on the even modes where the corresponding number of equations is $n_e$, the discussion for odd modes being entirely analogous.  Regularity of the wavefunctions at $\tau = 0$ fixes half of the $2n_e$ integration constants to zero.  We denote the remaining integration constants by $s_i$, $1 \leq i \leq n_e$.  For arbitrary $s_i$, evolving the differential equations to large $\tau$ generically yields non-normalizable wavefunctions. But because these equations are linear, the coefficients $c_i$ of the non-normalizable parts from \eqref{asymptotics} are linearly related to the $s_i$, and there exists some $n \times n$ matrix $\gamma(m^2)$ for which $c_i = \gamma_{ij} s_j$.  This matrix is degenerate precisely at those values of $m^2$ that solve the eigenvalue problem \eqref{EValue2}.

\begin{figure}%
\begin{center}
  \includegraphics[scale=1]{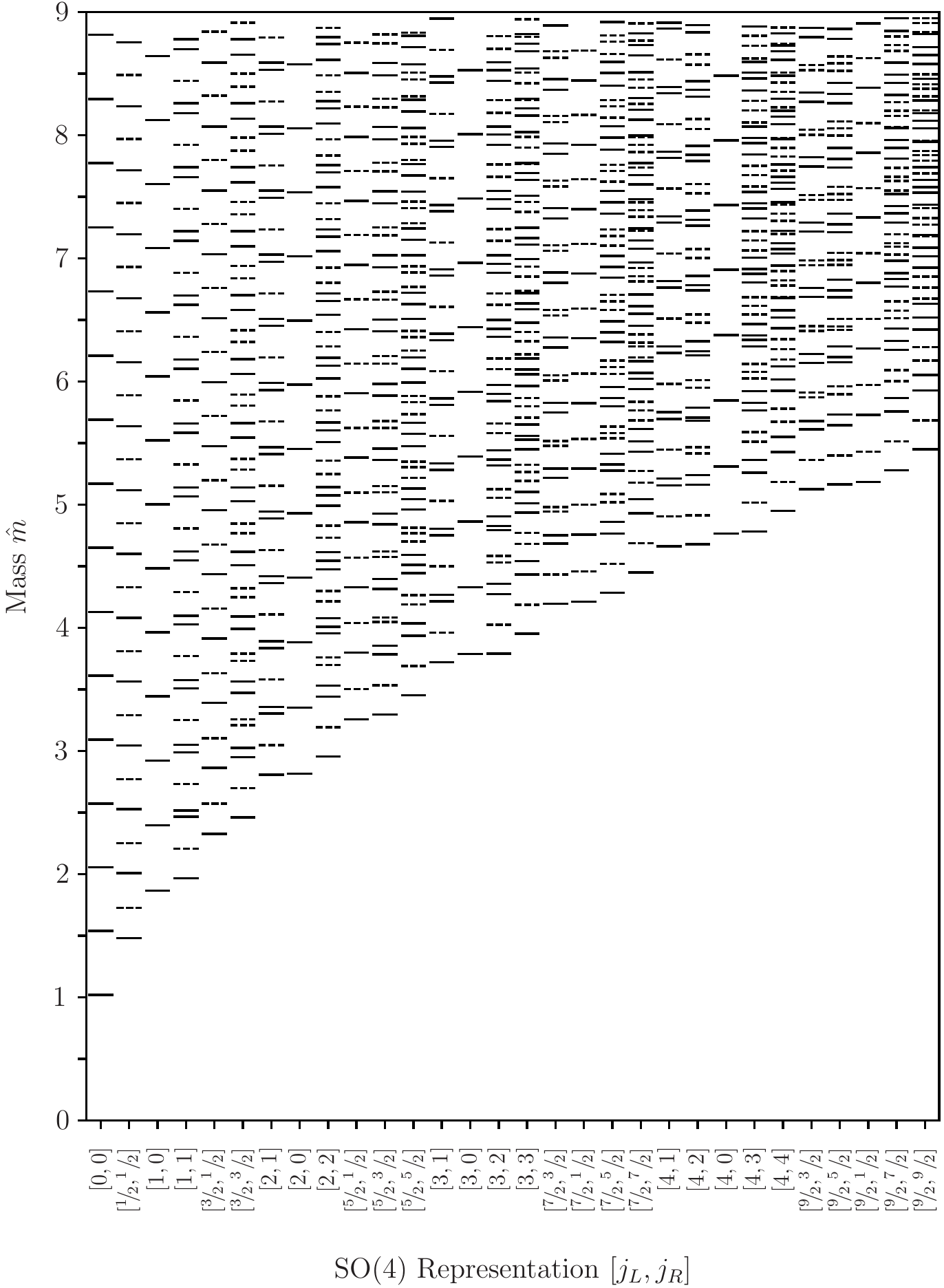}
\end{center}
\caption{Non-singlet spin-2 glueball masses for the warped deformed conifold \cite{Klebanov:2000hb}. Solid/dashed lines represent states with even/odd parity. The representations are ordered according to the mass of the lowest state. The numbers being plotted correspond to the values of $\hat m$ as defined in (\ref{mhatdef}).}%
\label{fig:MassesT11}%
\end{figure}

\begin{figure}%
\begin{center}
  \includegraphics[totalheight=0.35\textheight]{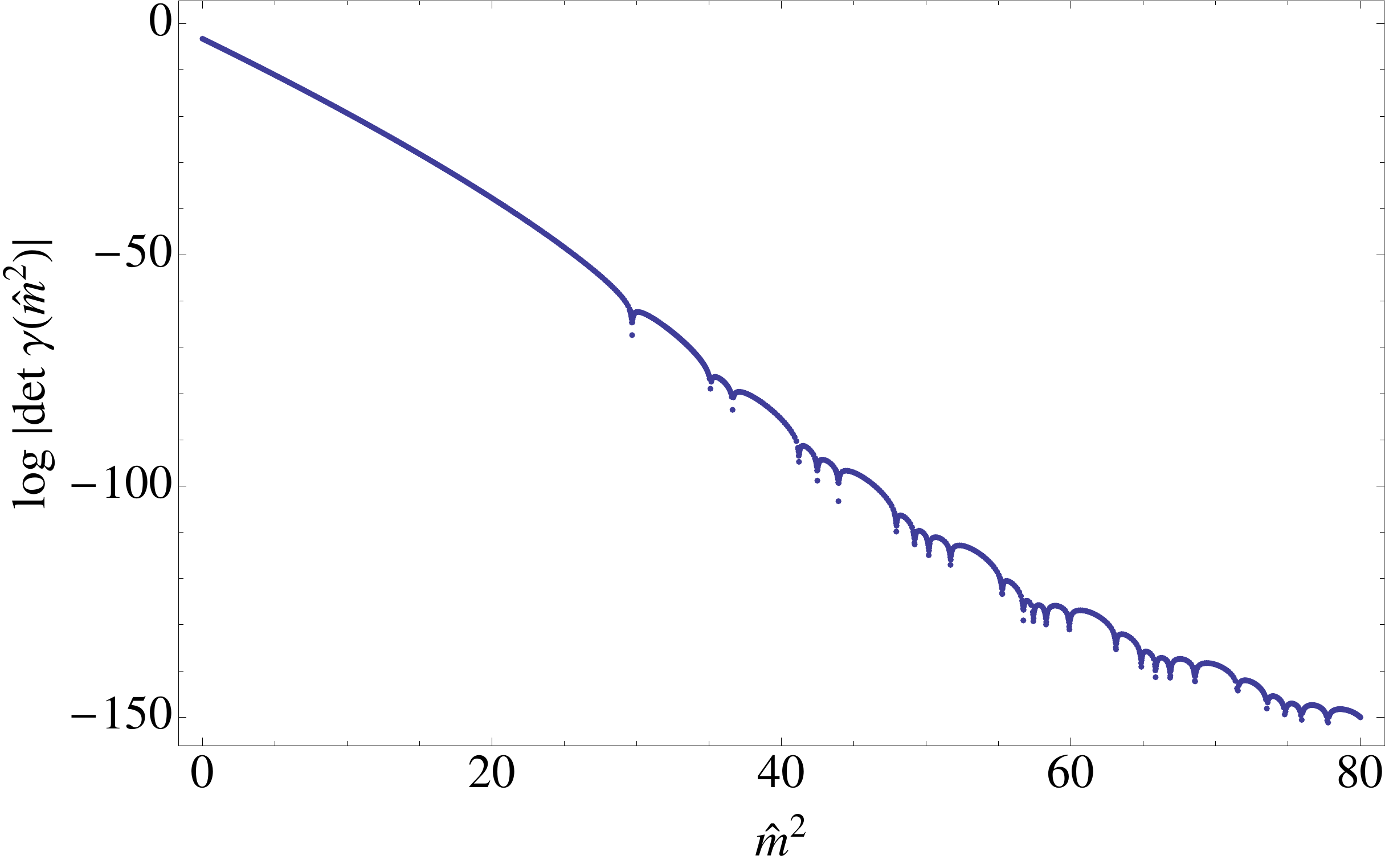}
\end{center}
\caption{The logarithm of $\abs{\det \gamma(m^2)}$ in the range $m^2 \in [0, 80]$ for the even sector of the $[\frac{9}{2},\frac{9}{2}]$ representation. The glueball masses are located at the dips in the plot. The masses can be seen to organize into 5 towers.}%
\label{fig:logdetplot}%
\end{figure}

To normalize the glueball masses, we define a dimensionless quantity $\hat m$, which agrees with the normalization of glueball masses in \cite{Berg:2006xy}, by
 \es{mhatdef}{
  m^2 = {2^{5/3} 3^{1/3} \over H(0) \epsilon^{4/3} } \hat m^2 \,,
 }
where $H(0) \approx 1.13983 (g_s M \alpha')^2 / \epsilon^{8/3}$.  We have computed the first fourteen values of $\hat m$ of each tower in all representations up to $[\frac{9}{2},\frac{9}{2}]$. The data is available online at \cite{GlueballWebsite}. See Figure~\ref{fig:MassesT11} for a summary of our results.
As an example of how one reads off the masses of various modes, we plotted the determinant of the matrix $\gamma$ described above for the even modes of the $[\frac{9}{2},\frac{9}{2}]$ representation in Figure \ref{fig:logdetplot}.

The masses within each tower exhibit a linear dependence on the excitation number $n$.  Furthermore, as was the case in \cite{Berg:2006xy}, the leading behavior appears to coincide for all modes.  This universal behavior at large $n$ has been derived using the WKB approximation \cite{Dymarsky:2007zs}.  In the following section we review their analysis.

\subsection{WKB Analysis}

The Laplacian acting on any eigenfunction $\phi$ can be written as
\es{wkbt11}{
  \left[\frac\partial{\partial\tau}(\sinh 2\tau - 2\tau)^{2/3}\frac\partial{\partial\tau} + \frac{G(\tau) \sinh^2\tau}{3\cdot 2^{1/3}}  +\frac{ m^2 H(\tau) \sinh^2\tau}{3\cdot 2^{1/3}} \right]\phi = 0 \; ,
}
where $G(\tau)$ is the angular contribution coming from $[g_{\cal C}(\tau) {\cal C} + g_{\cal R}(\tau) {\cal R} + g_{\cal L} (\tau) {\cal L}]\phi$, whose exact form will be irrelevant here. The rescaling $\phi(\tau) = (\sinh 2\tau - 2\tau)^{-1/3}\psi(\tau)$ then puts the equation in the Schr\"odinger-like form
\es{schrodinger}{[\partial^2_\tau - Q(\tau)]\psi = 0 \, ,}
where the effective potential is
\es{qt11}{
  Q(\tau) &= \frac{4}{3}\frac{\sinh 2\tau}{\sinh 2\tau - 2\tau}
    - \frac{32}{9}\frac{\sinh^4 \tau}{(\sinh 2\tau - 2\tau)^2}
    - \frac{\left[G(\tau) + m^2 H(\tau)\right] \sinh^2\tau}{3\cdot 2^{1/3}(\sinh 2\tau - 2\tau)^{2/3}}  \; .
 }
The Bohr-Sommerfeld quantization condition states that the glueball masses are given by solutions to
\es{BSquant}{\int_{\tau_0}^{\tau_1} \sqrt{-Q(x)dx} = n\pi - \frac{\pi}{2}}
where $n \in \mathbb{Z}^+$ and $\tau_i$ are the classical turning points at which $Q(\tau) =0$.\footnote{This equation assumes that the potential has sloping walls at both $\tau_0$ and $\tau_1$, which is correct if $j_L \neq j_R$. If $j_L = j_R$ there are modes for which there is no potential barrier at small $\tau$.  In this case it is standard to treat $\tau_0 = 0$ as a hard wall, and the RHS of \eqref{BSquant} becomes $n \pi - \frac{\pi}{4}$. } In the limit of very large $m$ and upon reinstating numerical factors to match \cite{Berg:2006xy}, we thus derive
\es{t11asymptotic}{
  m^2 \sim
   2^{1/3} 3 \pi^2
  \left(  \int_0^\infty \frac{\sqrt{H(\tau) \sinh\tau}}{(\sinh 2\tau - 2\tau)^{1/3}}\right)^{-2}n^2
  \quad \Rightarrow
  \quad \hat m^2 \approx 0.272\, n^2. }

\section{Generalization to Stenzel spaces}
\label{GENERALIZATION}

In section~\ref{GREEN} we explained how to reduce the calculation of the Green's function on the deformed conifold to solving coupled systems of differential equations.  In this section we generalize these results to the $d>4$ Stenzel spaces \eqref{DeformedConifold} (for the $d=3$ case, i.e.~the Eguchi-Hanson space \cite{Eguchi:1978xp}, there are further simplifications due to a symmetry enhancement from $\grSO(3)$ to $\grSO(3) \times \grU(1)$ and the Green's function is known in closed form \cite{Page:1979ga}).  In particular, the $d=5$ Stenzel space is relevant to the CGLP solution of the $11$-d supergravity equations of motion \cite{Cvetic:2000db} that is dual to a massive vacuum of a $(2+1)$-dimensional field theory.

\subsection{Ricci flat metric and laplacian}

The metric for arbitrary $d$ was already given in \eqref{MetricAgain}. In order to write down the Laplacian, we need to compute the determinant and the inverse of this metric. However, we cannot compute them directly from the expression \eqref{MetricAgain}, because not all coordinates $z_i$ and $\bar{z}_i$ ($i=1,\ldots,d$) are independent due to the complex constraint \eqref{DeformedConifold}. Instead, we first have to eliminate two coordinates, say $z_1$ and $\bar{z}_1$, from the metric using this constraint. Letting capital indices $A$ run over $(a,\bar{a})$ with $a=2,\ldots,d$, we can write the independent components of the metric as
\be
  ds^2 = G_{AB} dz_A dz_B \; ,
  \qquad
  G_{AB} = \matr{cc}{0 & G_{a\bar{b}} \\ G_{\bar{a}b} & 0} \; ,
\ee
where
\be
  G_{a\bar{b}} = \frac{\mathcal{F}'}{\epsilon^2 \sinh\tau}
                 \frac{ z_a \bar{z}_b + z_1 \bar{z}_1 \delta_{ab} }{2z_1\bar{z}_1}
               - \frac{\mathcal{F}''-\mathcal{F}'\coth\tau}{\epsilon^4 \sinh^2\tau}
                 \frac{\brk{ \bar{z}_1 z_a - \bar{z}_a z_1 } \bigbrk{ \bar{z}_1 z_b - \bar{z}_b z_1 }}{2z_1\bar{z}_1}
\ee
and $G_{\bar{a}b} \equiv G_{b\bar{a}}$. Now, it is straightforward to compute the determinant\footnote{\label{ftn:ricciflat} In passing we note that one can now see that the Ricci flatness condition $R_{a\bar{b}} = \partial_a \bar{\partial}_b \ln \sqrt{\det G} = 0$ for \eqref{eqn:det} leads to an equation like \eqref{eqn:ricciflat} with some constant right hand side. \TK{What made us choose this constant?} The purely holomorphic and anti-holomorphic factors $z_1$ and $\bar{z}_1$ vanish when taking the logarithm and the mixed derivatives.}
\be \label{eqn:det}
  G = \det G_{AB} = -(-1)^d \lrbrk{ \frac{\mathcal{F}'}{2\epsilon^2\sinh\tau} }^{2(d-2)}
                                  \lrbrk{ \frac{\mathcal{F}''}{2z_1\bar{z}_1} }^2
\ee
and the inverse metric
\be
  G^{AB} = \matr{cc}{0 & G^{a\bar{b}} \\ G^{\bar{a}b} & 0}
\ee
where
\be
  G^{a\bar{b}} \eq \frac{2\epsilon^2\sinh\tau}{\mathcal{F}'} \delta_{ab}
               + \frac{2}{\mathcal{F}'' } \Biggsbrk{
                  - \bar{z}_a z_b
                  + \frac{\mathcal{F}''-\mathcal{F}'\coth\tau}{\mathcal{F}' \epsilon^2 \sinh\tau} \sum_{i=1}^{d} ( \bar{z}_i z_a - \bar{z}_a z_i ) ( \bar{z}_i z_b - \bar{z}_b z_i )
               }
\ee
and $G^{\bar{a}b} = G^{b\bar{a}}$.

In terms of these functions, the Laplacian is given by
\be \label{eqn:lapl-actual}
  \lap_{2d-2} \eq \frac{1}{\sqrt{G}} \frac\partial{\partial z_A} \Bigbrk{ \sqrt{G} G^{AB} \frac\partial{\partial z_B} }
            = \frac{1}{\sqrt{G}} \frac\partial{\partial z_a} \Bigbrk{ \sqrt{G} G^{a\bar{b}} \frac\partial{\partial \bar{z}_b}} + \mathrm{c.c.}
\ee
where in this formula, $z_1 = \sqrt{\eps^2 - z_a z_a}$ and $\bar{z}_1 = \sqrt{\eps^2 - \bar{z}_a \bar{z}_a}$ are understood to be functions of the other variables. In order to write the Laplacian in a manifestly $\grO(d)$ invariant form, we replace
\be \label{eqn:replacement}
  \frac{\partial}{\partial z_a} \rightarrow \frac{\partial}{\partial z_a} - \frac{z_a}{z_1} \frac{\partial}{\partial z_1}
  \; , \qquad
  \frac\partial{\partial\bar{z}_a} \rightarrow \frac{\partial}{\partial\bar{z}_a} - \frac{\bar{z}_a}{\bar{z}_1} \frac{\partial}{\partial\bar{z}_1}
\ee
and treat $z_1$ and $\bar{z}_1$ as independent from the other coordinates. The reason for \eqref{eqn:replacement} is the following. We want to act on functions $f(z_i,\bar{z}_j)$ where so far $z_1$ and $\bar{z}_1$ were dependent on $z_A$ and $\bar{z}_A$. The derivative $\partial/\partial z_a$ is thus
\be
  \frac{\partial f}{\partial z_a} + \frac{\partial z_1}{\partial z_a} \frac{\partial f}{\partial z_1} \; ,
\ee
and similarly for $\partial/{\partial \bar{z}_a}$. With $(\partial/\partial z_a) z_1 = - z_a/z_1$ we find \eqref{eqn:replacement}. Making these replacements, \eqref{eqn:lapl-actual} becomes
\be \label{eqn:lapl-z}
  \lap_{2d-2} \eq \frac{4}{\mathcal{F}''}
  \biggsbrk{
    \Half \left(z_i \frac{\partial}{\partial z_j} - z_j \frac\partial{\partial z_i}\right) \left(\bar{z}_i \frac\partial{\partial\bar{z}_j} - \bar{z}_j \frac\partial{\partial\bar{z}_i}\right) \\ && \hspace{0mm}
    +\frac{\mathcal{F}''-\mathcal{F}'\coth\tau}{\mathcal{F}' \epsilon^2 \sinh\tau}\Bigbrk{
      \epsilon^4\sinh^2\tau \delta_{ij} - \epsilon^2\cosh\tau (z_i \bar{z}_j + \bar{z}_i z_j) + \epsilon^2 (z_i z_j + \bar{z}_i \bar{z}_j)
    } \frac{\partial}{\partial z_i} \frac\partial{\partial \bar{z}_j}
  } \; . \nn
\ee
It is obvious that this operator annihilates the constraint \eqref{DeformedConifold} as every term contains one holomorphic and one anti-holomorphic derivative. The fact that \eqref{eqn:lapl-z} annihilates the constraints is a prerequisite for this equation to make sense.

To bring the Laplacian into a form that corresponds to \eqref{LapDeformedT11} in the $d=4$ case, we pass to a ``radial'' coordinate $\tau$ and a set of ``angles'' $(y_i, \bar{y}_j)$ that parametrize the base of the undeformed conifold.  These coordinates are defined in \eqref{zToy}.  Since the $y$'s are even more constrained than the $z$'s, it is easiest to work backwards, i.e.\ to guess an expression for the Laplacian in terms of $\tau$ and $y$ and then to verify the guess by changing coordinates back to $z$.  Inspired by the Laplacian $\lap_6$ on the conifold ($d=4$), we find by trial and error:
\es{eqn:lapl-y}{
  \lap_{2d-2} &= \frac{4}{\mathcal{F}''} \Biggl[
    (d-2) \coth\tau \, \partial_\tau
    +  \partial_\tau^2
    - \Half \coth^2\tau \, {\cal C}
    + \Quarter (1 + 2\csch^2\tau) {\cal R}
    + \frac{\cosh\tau}{\sinh^2\tau} {\cal L} \cr
   {}&+\frac{\mathcal{F}''-\mathcal{F}'\coth\tau}{\mathcal{F}' \epsilon^2 \sinh\tau}\biggl(
    (d-2) \eps^2\sinh\tau\partial_\tau
    - \Half \epsilon^2\cosh\tau \, {\cal C}
    + \Half \epsilon^2\cosh\tau \, {\cal R}
    + \epsilon^2 {\cal L} \biggr) \Biggr]
  }
where
\es{DiffOps}{
  {\cal C}    &= y_i y_j \frac{\partial^2}{\partial{y_i}\partial{y_j}}
                    + (\bar{y}_i y_j - \delta_{ij} y_k \bar{y}_k) \frac{\partial^2}{\partial{y_i} \partial \bar{y}_j}
                    + (d-1) y_i \frac\partial{\partial y_i} + \mathrm{c.c.} \; , \\
  {\cal R} &= \biggbrk{ y_i \frac\partial{\partial{y_i}} - \bar{y}_i \frac\partial{\partial\bar{y}_i} }
                      \biggbrk{ y_j \frac\partial{\partial{y_j}} - \bar{y}_j \frac\partial{\partial\bar{y}_j} } \; , \\
  {\cal L}    &= \frac 1 2 \bigbrk{ \bar{y}_i y_j + y_i \bar{y}_j - \delta_{ij} y_k \bar{y}_k } \frac{\partial^2}{\partial y_i\partial y_j}
                    + \frac{d-2}{2} \bar{y}_i \frac\partial{\partial y_i} + \mathrm{c.c.}
}
Here ${\cal C}$ is the Casimir operator of $\grSO(d)$. Note also that ${\cal C}$, ${\cal R}$, and ${\cal L}$ annihilate the constraints \eqref{yConstraints} or result in expressions that vanish when the constraints are used. In order to verify that \eqref{eqn:lapl-y} is equal to \eqref{eqn:lapl-z}, we need the inverse of \eqref{zToy} which is given by
\be
  \tau      \eq \arccosh(z_i \bar{z}_i / \epsilon^2) \label{eqn:tauofz} \; , \\
  y_i       \eq \frac{\epsilon}{\sqrt{2}} \frac{e^{\tau/2} z_i       - e^{-\tau/2} \bar{z}_i}{\epsilon^2 \sinh\tau} \; ,
\ee
where the $\tau$ in the second equation is just a shorthand for the right hand side of the first equation.  We also need expressions for the $y$- and $\tau$-derivatives in terms of the $z$-derivatives. These are found from \eqref{zToy} by the chain rule:
\be
  \frac\partial{\partial y_i} \eq \frac{\epsilon}{\sqrt{2}}
                                  \biggbrk{ e^{\tau/2}  \frac\partial{\partial z_i}
                                          + e^{-\tau/2} \frac\partial{\partial\bar{z}_i} } \;  , \label{eqn:dyTodz} \\
  \partial_{\tau}             \eq \Half \coth\tau \biggbrk{ z_i \frac\partial{\partial {z_i}} + \bar{z}_i \frac\partial{\partial \bar{z}_i} }
                                - \Half \csch\tau \biggbrk{ z_i \frac\partial{\partial \bar{z}_i} + \bar{z}_i \frac\partial{\partial z_i} } \; ,
\ee
where again $\tau$ is to be substituted by \eqref{eqn:tauofz}. This substitution is important when computing the second order derivatives.

Using the functions $g_{\cal C}(\tau)$, $g_{\cal R}(\tau)$, and $g_{\cal L}(\tau)$ defined in \eqref{Gotfs}, the Laplacian can be brought into the concise form that we had in the $d=4$ case:
\be \label{lapGeneral}
  \lap_{2d-2} \eq
    {\cal T} + g_{\cal C}(\tau) {\cal C} + g_{\cal R}(\tau) {\cal R} + g_{\cal L}(\tau) {\cal L}
\ee
with
\be \label{LtauGeneral}
  {\cal T} = \frac{4}{\mathcal{F}'' \mathcal{F}'^{d-2}} \partial_\tau \bigbrk{ \mathcal{F}'^{d-2} \partial_\tau } \; ,
\ee
and the other operators as in \eqref{DiffOps}. When plugging the $d=4$ expression \eqref{FpT11} for $\mathcal{F}'$ into ${\cal T}$, we recover \eqref{GotLrho}.

\subsection{Functions on Stenzel spaces}

As in the $d=4$ case, functions on the Stenzel spaces may be expanded in homogeneous polynomials $F(y_i,\bar{y}_i)$---see eq.~\eqref{GeneralWavefunction}---with $\tau$-dependent coefficients. The tensor
\be \label{eqn:tensor}
  M_{i_1 i_2 \cdots i_{n_1}}^{j_1 j_2 \cdots j_{n_2}} \; ,
\ee
in \eqref{Polyn} now transforms in a representation of $\grSO(d)$ but otherwise has the same properties as before. We want to find its irreducible components.

We denote $\grSO(d)$ (vector) representations by their Young tableaux $(p_1,p_2,\ldots,p_r)$, where $p_i$ is the number of boxes in row $i$ and the total number of rows equals the rank $r=\bigfloor{d/2}$ of $\grSO(d)$. The fundamental representation $\rep{d}$ is $(1,0,\ldots,0)$. The $k$-index symmetric-traceless representation is $(k,0,\ldots,0)$.

Group theoretically the tensor \eqref{eqn:tensor} is a direct product of the two irreps $(n_1,0,\dots,0)$ and $(n_2,0,\dots,0)$, which represent the two sets of symmetric-traceless lower and upper indices. This product decomposes into the following representations \cite{Girardi:1981jz,Girardi:1981ss} (where $n_1 \ge n_2$, otherwise multiply the representations in the other order):
\be
 (n_1,0,\ldots,0) \otimes (n_2,0,\ldots,0) = \sum_{j=0}^{n_2} \sum_{i} (n_1+n_2-i-2j,i,0,\dots,0) \; ,
\ee
where $i$ satisfies $n_1+n_2-2j \ge 2i \ge 0$ and $n_2-j\ge i$. The representations on the right hand side differ by the number of anti-symmetrizations (counted by the index $i$) and the number of traces (counted by the index $j$) between lower and upper indices. For instance, a representation with $j=1$ will contain tensors that are proportional to one delta function $M_{i_1 \cdots i_{n_1}}^{j_1 \cdots j_{n_2}} \sim \delta^{j_n}_{i_m}$. However, we do not want to include any such representation, because the corresponding homogeneous polynomial would reduce to one with fewer powers of $y$ and $\bar{y}$ due to the constraint $\sum_{i=1}^d y_i \bar{y}_i = 1$. Therefore, the only representations that we need to consider are those with $j=0$:
\be \label{FinalDecomp-SOd}
 \Bigsbrk{ (n_1,0,\ldots,0) \otimes (n_2,0,\ldots,0) }_{\mathrm{traceless}} = \sum_{i=0}^{\min(n_1,n_2)} (n_1+n_2-i,i,0,\dots,0)
\ee
This sum of representations corresponds to \eqref{FinalDecomp} in the $d=4$ case.

Thus, we need to understand representations of the form
\be \label{eqn:rep-p-q}
  \includegraphics{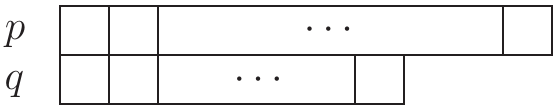}
\ee
alias $(p,q,0,\ldots,0)$. Their dimensions are
\be
  D_{d,p,q} \eq \frac{(p-q+1)(p+q+d-3)(2p+d-2)(2q+d-4)}{(d-2)(d-3)(d-4)^2} \times \nl
          \times \binom{p+d-4}{p+1} \binom{q+d-5}{q}
\ee
and the eigenvalue of the quadratic Casimir in \eqref{DiffOps} is given by
\be
  {\cal C} = p(p+d-2) + q(q+d-4) \; .
\ee

Just as in the $\grSO(4)$ case, the representation \eqref{FinalDecomp-SOd} for some given $p$ and $q$ can be realized by different choices for $n_1$ and $n_2$ as long as $p+q = n_1+n_2$.  There are a total of $p-q+1$ copies of the same $(p, q, 0, \ldots, 0)$ irrep of $\grSO(d)$, corresponding to the spin-${p-q\over 2}$ representation of $\widetilde{\grSU(2)}$.  The highest weight state of this $\widetilde{\grSU(2)}$ irrep is the polynomial with the largest number of $y$'s obtained for $n_1=p$ and $n_2=q$. In the case $d = 4$, this polynomial was given in eq.~\eqref{HighestWeight} with $n_1=p=j_L + j_R$ and $n_2=q=\abs{j_L - j_R}$, but it can be generalized to arbitrary $d$ by writing
\be
  h_{p,q} = (y_2 + i y_3)^{p-q} \bigsbrk{ (y_2 + i y_3)(\bar{y}_1 + i \bar{y}_4) - (y_1 + i y_4)(\bar{y}_2 + i \bar{y}_3) }^q \; .
\ee
Acting repeatedly with $\tilde J_-$ we obtain
\be
  F_{p,q,\tilde{m}} \eq \sqrt{ { p-q \choose \frac{p-q}{2} - \tilde{m}} }
                      \, (y_2 + i y_3)^{\frac{p-q}{2}+\tilde{m}}
                      \, (\bar{y}_2 + i \bar{y}_3)^{\frac{p-q}{2}-\tilde{m}} \nl
                      \hspace{13mm} \times \bigsbrk{ (y_2 + i y_3)(\bar{y}_1 + i \bar{y}_4) - (y_1 + i y_4)(\bar{y}_2 + i \bar{y}_3) }^q  \; .
\ee

\subsection{Green's function computation}

We can now present a generalization of the Green's function computation from section~\ref{GREEN} to arbitrary $d$.  Suppose we want to solve the Laplace equation with a delta-function source at $\vec u_0 = (1, 0, \ldots, 0)$,
 \es{GeqGeneral}{
  \lap_{2d-2} G(\tau, y_i; \vec{u}_0) =
    \frac{d-1}{d-2} \, \frac{1}{\epsilon^{2(d-2)}} \,
    \delta^{(d-1)} (\vec{u} - \vec{u}_0)
    \delta^{(d-1)} (\vec{w}) \; .
 }
Here, we used the decomposition $\vec{y} = \vec{u} + i \vec{v}$ as in \eqref{yDef} and denoted $\vec{w} = {\tau \over 2} \vec{v}$ as in section~\ref{GREEN}.  The right hand side is normalized such that it integrates to unity over the generalized deformed conifold.  The Green's function that solves this equation can be expanded in terms of polynomials in $y_i$ and $\bar y_i$ with $\tau$-dependent coefficients.  Since the source preserves an $\grSO(d-1)$ symmetry that rotates the last $d-1$ $y_i$ into each other, the polynomials that appear in the expansion of the Green's function also have to be invariant under this $\grSO(d-1)$.  These polynomials are just functions of $y_1$ and $\bar y_1$ and consequently they transform in $\grSO(d)$ irreps with $q = 0$.  For the representation with highest weight $(p, 0, 0, \ldots)$, there are $p+1$ such polynomials indexed by the label $\tilde m$ ranging between $\pm p/2$.

One can find the form of the polynomials that appear in the expansion of the Green's function by making the ansatz $F(y_1, \bar y_1) = y_1^{2 \tilde m} f(y_1 \bar y_1)$ or $F(y_1, \bar y_1) = \bar y_1^{2 \tilde m} f(y_1 \bar y_1)$ and solving the equation ${\cal C} F = p(p+d-2) F$.  The resulting functions are
 \es{GotFsGenerald}{
   F_{p\tilde m}(y_1, \bar y_1) = {\cal N}_{p\tilde m}\, {}_2 F_1 \left(\abs{\tilde m}-{p \over 2}, {p+d \over 2} + \abs{\tilde m}-1, 2\abs{\tilde m}+1; 2 y_1 \bar y_1 \right)  \times \begin{cases}
   y_1^{2 \tilde m} & \text{if $\tilde m \geq 0$} \; , \\
   \bar y_1^{2 \abs{\tilde m}} & \text{if $\tilde m < 0$} \; .
   \end{cases}
 }
The hypergeometric function in \eqref{GotFsGenerald} reduces to a polynomials of degree ${p\over 2} - \abs{\tilde m}$ in $y_1 \bar y_1$.  If we choose the normalization constant to be \TK{Should we include another factor of $\sqrt{p+1}$ in order to match \eqref{FNorm}?}
 \es{NormConstGenerald}{
  {\cal N}_{p\tilde m} = (-1)^{-\frac{p}{2}} \, (-2)^{\abs{\tilde m}} \,
   { \left( {d +p \over 2} + \abs{\tilde m} -2 \right) !
   \over (2 \abs{\tilde m})! \left({p\over 2} - \abs{\tilde m} \right)!} \; .
 }
we then have
 \es{OpsGenerald}{
  {\cal C} F_{p\tilde m} &= p(p+d-2) F_{p\tilde m} \; , \\
  {\cal R} F_{p\tilde m} &= 4\tilde m^2 F_{p\tilde m} \; , \\
  {\cal L} F_{p\tilde m} &= \Bigbrk{ {p\over 2} - \tilde m + 1 } \Bigbrk{ {p+d \over 2} + \tilde m -2 } F_{p, \tilde m -1} \\
   {}&+ \Bigbrk{ {p \over 2} + \tilde m + 1 } \Bigbrk{ {p + d \over 2} - \tilde m - 2 } F_{p, \tilde m + 1} \; .
 }

The Green's function solving \eqref{GeqGeneral} can thus be expanded as
 \es{GreensFunctionGeneral}{
  G(\tau, y_i; \vec{u}_0) = \sum_{p = 0}^\infty \sum_{\tilde m = -{p\over 2}}^{p\over 2} f_{p\tilde m}(\tau) F_{p \tilde m}(y_1, \bar y_1) \; ,
 }
for some coefficient functions $f_{p\tilde m}(\tau)$.  Using eqs.~\eqref{Gotfs}, \eqref{lapGeneral}, \eqref{LtauGeneral}, and \eqref{OpsGenerald} it is straightforward to find the differential equations for the $f_{p\tilde m} (\tau)$.  Since the Laplacian on the deformed conifold is an $\grSO(d)$ singlet operator, it only mixes together functions with the same $p$.  The $\Z_2$ symmetry of the deformed conifold and the source further requires $f_{p\tilde m}(\tau) = f_{p,-\tilde m}(\tau)$, so for each $p$ we have to solve for $n_p = \bigfloor{\frac{p}{2}}+1$ independent functions.  As in the $d=4$ case we will not find these functions explicitly, but instead will give a prescription for finding them.

For each $p$, there are $n_p = \bigfloor{\frac{p}{2}}+1$ second order ODEs which give a total of $2n_p$ integration constants.  Of these, $n_p$ are fixed by requiring $f_{p\tilde m}$ to vanish at large $\tau$.  The remaining $n_p$ integration constants should be fixed by imposing appropriate boundary conditions at $\tau = 0$.  Close to $\tau =0$, the differential equation \eqref{GeqGeneral} takes the approximate form
 \be \label{GreenCloseGeneral}
    && \lrsbrk{\frac{1}{\tau^{d-2}} \frac{\partial}{\partial \tau} \tau^{d-2} \frac{\partial}{\partial\tau}
    + \frac{-{\cal C} + {\cal R} + 2 {\cal L} }{2 \tau^2} } G(\tau, y_i; \vec{u}_0)
    = \nln
    && \hspace{30mm} = \frac{1}{4} \lrbrk{ \frac{d-1}{d-2} \, \frac{1}{\epsilon^{2(d-2)}} }^{\frac{d-2}{d-1}} \,
      \delta^{(d-1)}(\vec{u} - \vec{u}_0) \delta^{(d-1)} (\vec{w}) \; .
 \ee
By arguments similar to those in section~\ref{GREEN}, at small $\tau$ we require $f_{p\tilde m}$ to behave as
 \es{SmalltauGeneral}{
  f_{p\tilde m} = {c_{p \tilde m} \over \tau^{d-3}} + \order\Bigbrk{\frac{1}{\tau^{d-4}}}
 }
with the constants $c_{p\tilde m}$ chosen in such a way as to reproduce the delta-function structure in \eqref{GreenCloseGeneral}.  For a given $p$, imposing \eqref{SmalltauGeneral} requires the $c_{p\tilde m}$ to be such that $\sum_{\tilde m} c_{p\tilde m} F_{p\tilde m}$ is annihilated by the operator $-{\cal C} + {\cal R} + 2 {\cal L}$.  From \eqref{OpsGenerald}, it can be seen that the kernel of this operator is spanned by
 \es{EpDefGenerald}{
  E_p = \sum_{\tilde m = -{p\over 2}}^{p \over 2}  F_{p\tilde m} \; ,
 }
so $c_{p\tilde m} = c_p$ is independent of $\tilde{m}$.  More insight into the $E_p$ can be obtained by writing $-{\cal C} + {\cal R} + 2 {\cal L}$ as a differential operator
 \es{DiffOpuv}{
  -{\cal C} + {\cal R} + 2 {\cal L}
   = -2(d-2) v_1 {\partial \over \partial v_1} + 2(1 - u_1^2 - v_1^2) {\partial^2 \over \partial v_1^2}
 }
when acting on functions of only $u_1$ and $v_1$.  The only normalizable functions of $u_1$ and $v_1$ that are annihilated by this operator are those that are independent of $v_1$, so $E_p$ indeed only depends on $u_1$.  Using the explicit formulae for $F_{p\tilde m}$ in \eqref{GotFsGenerald}, one can write
\be
  E_p = \Gamma\biggbrk{\frac{d-2}{2}} \, C_p^{d-2\over 2} (\cos \theta) \; ,
\ee
where $u_1 = \cos\theta$ and $C_n^\alpha(x)$ are the ultraspherical (Gegenbauer) polynomials.\footnote{Since $E_p$ is only a function of $u_1$ and transforms in $(p, 0, \ldots)$ of $\grSO(d)$, it must be proportional to one of the zonal spherical harmonics $Z_p(\theta)$.  The zonal spherical harmonics are
\[Z_p (\theta) = \sqrt{ \frac{ 2 p + d - 2 }{ d-2 } \, \frac{ p! \, (d-3)! }{ (p + d - 3)! } } \, C_p^{d-2\over 2} (\cos \theta)\; ,
 \qquad {1 \over \Vol(S^{d-1})} \int d \Omega_{S^{d-1}} \, Z_p(\theta) Z_{p'}^*(\theta)  = \delta_{pp'} \; ,\]
where $d\Omega_{S^{d-1}}$ is the standard volume element on $S^{d-1}$.
} Choosing
 \es{cpChoice}{
  c_{p \tilde m} = c_p = - \frac{1}{32\pi^{d-1}} \, \frac{2p+d-2}{(d-1)(d-3)} \, \frac{\Gamma(d+1)}{\Gamma\!\lrbrk{\frac{d+2}{2}}} \, \lrbrk{ \frac{d-1}{d-2} \, \frac{1}{\epsilon^{2(d-2)}} }^{\frac{d-2}{d-1}} \; ,
 }
one can check that the completeness relation
 \es{CompletenessGenerald}{
  \delta^{(d-1)}(\vec{u} - \vec{u}_0) = {1 \over \Vol(S^{d-1})} \sum_p {2p + d - 2\over d-2} \, C_p^{d-2\over 2}(\cos\theta)
 }
implies that the LHS of \eqref{GreenCloseGeneral} reproduces the delta-function structure of the RHS of the same equation.

\section{Bound state masses for the CGLP background}
\label{GLUEBALLSM}

\begin{figure}%
\begin{center}
  \includegraphics[scale=1]{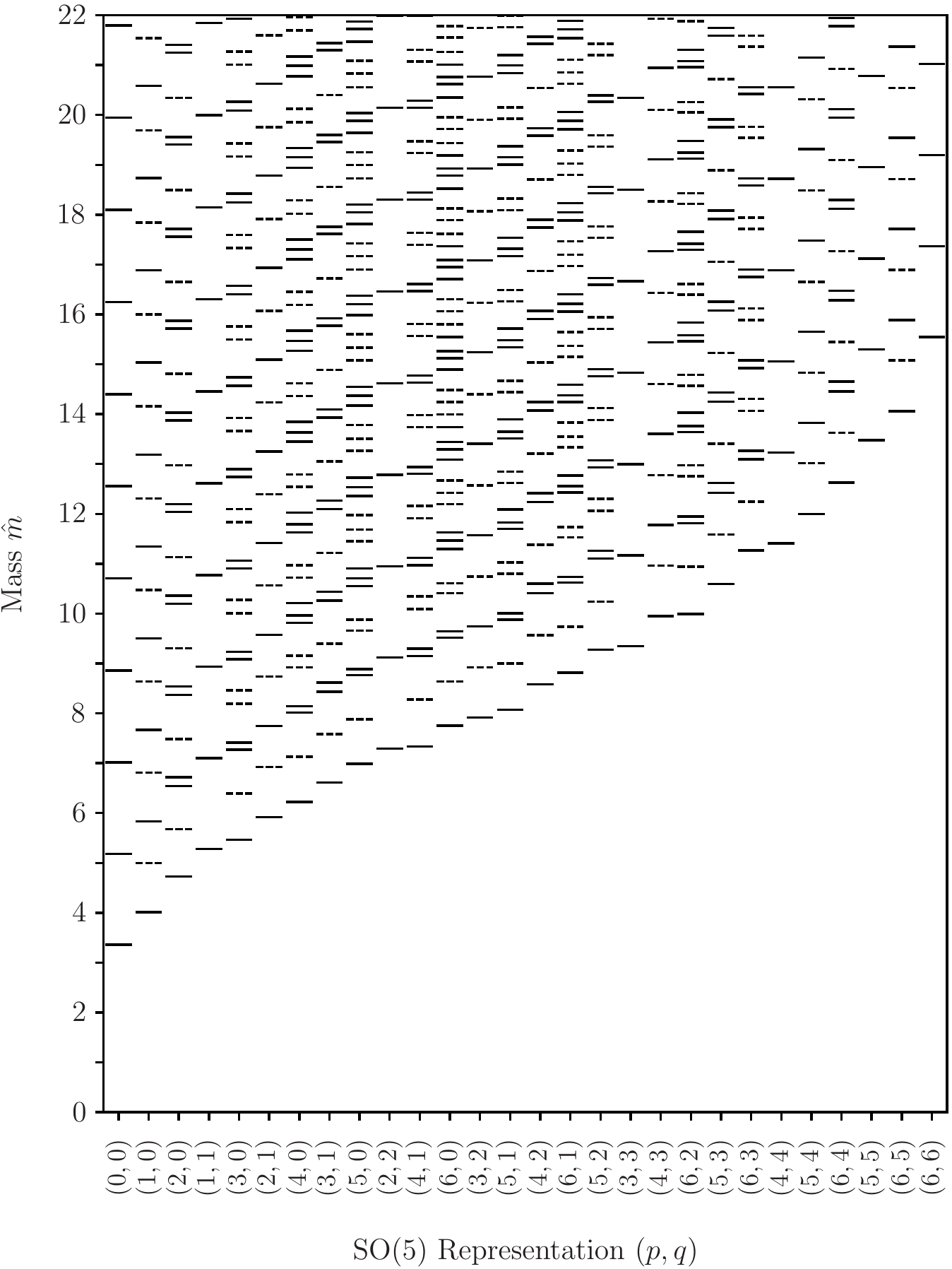}
\end{center}
\caption{Glueball masses for spin-2 excitations of the warped M-theory background \cite{Cvetic:2000db} constructed from the 8-d Stenzel space.  Solid/dashed lines represent states with even/odd parity. The representations are ordered according to the mass of the lowest state.  The numbers being plotted correspond to $\hat m$ as defined in (\ref{mhatDefM}).}%
\label{fig:MassesV52}%
\end{figure}

The CGLP background \cite{Cvetic:2000db}, which is an M-theory generalization of the KS background, has the metric
\es{MTheoryMetric}{
  ds^2 = H^{-\frac 2 3}dx_\mu dx^\mu + H^{\frac 1 3}ds^2_8 \; ,
}
where $dx_\mu dx^\mu$ here denotes the Minkowski metric on $\mathbb{R}^{2,1}$, $ds_8^2$ is the metric on the $d=5$ Stenzel space, and the warp factor is given by
\es{mwarp}{
  H(\tau) = \frac{3^{\frac{11} 4}}{\sqrt{2}} {m_{\rm CGLP}^2 \over \epsilon^{9\over2}}  \int_{\tau}^{\infty} dx\, \frac{\sinh x}{(2+\cosh x)^{3/4}(1+\cosh x)^{5/2}} \; ,
}
where $m_{\rm CGLP}$ is the quantity denoted by $m$ in \cite{Klebanov:2010qs}.  As before, we find the glueball masses by solving an eigenvalue problem, namely
\es{EValue3}{\lap_8\phi = -m^2H(\tau)\phi.}  The decomposition of $\lap_8$ outlined in eqs.~\eqref{eqn:lapl-y}--\eqref{DiffOps} reduces the problem to a finite system of coupled ODEs. Furthermore, the asymptotic behavior of the eigenmodes in the UV is known \cite{Ceresole:1999zg}. The identification of eigenmodes thus proceeds exactly by analogy to Section~\ref{GLUEBALLSIIB}.  To normalize the glueball masses, we defined the dimensionless mass
 \es{mhatDefM}{
  \hat m = m {m_{\rm CGLP} \over \epsilon^{3/2} } \; .
 }
We plot the spectrum of glueball masses for the representations $(p,q)=(0,0)$ to $(6,6)$ in Figure~\ref{fig:MassesV52}. The numerical values are available at \cite{GlueballWebsite}.

Here too, $m^2$ exhibits a quadratic dependence on excitation number. To demonstrate this dependence analytically, we again write down the generic Laplacian in this geometry,
\es{LaplV52}{
  \left[\frac \partial{\partial\tau}(2+\cosh \tau)^{3/4}\sinh^3\frac{\tau}{2}\frac\partial{\partial\tau} + \frac{3^{7/4}\sinh^3\frac{\tau}{2}\cosh^3\frac{\tau}{2}[G(\tau) + m^2H(\tau)]}{8}\right] \phi = 0 \; ,
}
where $G(\tau)$ represents the angular contribution. Then the field redefinition $\phi(\tau) = (2 + \cosh\tau)^{-3/8}(\sinh\tau/2)^{-3/2}\psi(\tau)$ puts it in the Schr\"odinger-like form \eqref{schrodinger} where
\es{QV52}{
 Q(\tau) = \frac{27}{128}\frac{(-7+12\cosh\tau + 3 \cosh 2\tau)\coth^2\frac{\tau}{2}}{(2+\cosh\tau)^2} - \frac{3^{7/4}}{8}\frac{[G(\tau)+m^2H(\tau)]\cosh^3\frac{\tau}{2}}{(2+\cosh \tau)^{3/4}}  \;.
}
In the large $n$ limit, it can thus be seen that
\es{v52asymptotic}{
  m^2 \sim \frac{8\pi^2}{3^{7/4}}\left(\int_0^\infty\frac{(\cosh\frac \tau 2)^{3/2}\sqrt{H(\tau)}}{(2+\cosh \tau)^{3/8}}\right)^{-2} n^2
  \quad \Rightarrow \quad \hat m^2 \approx 3.433n^2 \; .
}

\section{Discussion}

In this paper we have calculated certain Green's functions and normal mode spectra on (warped) Stenzel spaces, and discussed the implications of our results for gauge/gravity duality.  The Green functions determine the back-reaction of an additional stack of branes placed on a warped Stenzel space. Their calculation necessitates solving a mixing problem for modes that have the same $\grSO(d)$ quantum numbers but different asymptotic $\grU(1)_R$ charges. Such a problem can only be solved numerically, but we have given a detailed prescription for such a calculation, at least in the $\grSO(d-1)$-symmetric case where the D3- or M2-branes are placed at the tip of the deformed cone.

An interesting problem for further work is to replace the branes by anti-branes. This provides a description of metastable states in the gauge theories dual to warped Stenzel spaces \cite{Kachru:2002gs,Klebanov:2010qs}. In this context, it is interesting to study the back-reaction of the stack of anti-D3 or anti-M2 branes. For anti-D3 branes on the KS background this was considered in \cite{Bena:2009xk}, but only when the branes were smeared. Perhaps our calculation of the Green function will eventually help attack the more interesting case of localized anti-D3 branes.

We have also addressed the eigenvalue problems that arise in the calculations of bound state (glueball) spectra transforming in the non-singlet representations of $\grSO(d)$. So far, we have limited ourselves to the spectrum of the minimal scalar equation, which corresponds to longitudinally polarized gravitons. It is of obvious interest to extend the glueball calculations to other non-singlet modes.

\section*{Acknowledgments}
We thank A.~Dymarsky, N.~Kamburov, S.~Kuperstein, S.~Sivek, and T.~Tesileanu for useful discussions.  This work was supported in part by the US NSF  under Grant No. PHY-0756966. SSP was also partly supported by a Princeton University Porter Ogden Jacobus Fellowship.


\appendix


\section{Metric on Stenzel spaces}
\label{V52METRIC}

In an earlier paper by Cvetic, Gibbons, L\"u, and Pope \cite{Cvetic:2000db}, the metric on the deformed conifold, \eqref{dsy}, was written in terms of $\grSO(d)/\grSO(d-2)$ rotation generators. The $\grSO(d)$ rotation generators $L_{AB}$, $A=1,\ldots,d$ were denoted by
\be
  \nu = L_{12}
  \; , \qquad
  \sigma_i = L_{1i}
  \; , \qquad
  \tilde{\sigma}_i = L_{2i}
\ee
where $i = 3,\ldots,d$, and the metric was written as
\be \label{cglp}
  ds^2 = \quarter c^2 d\tau^2 + a^2 \sigma_i^2 + b^2 \tilde{\sigma}_i^2 + c^2 \nu^2
\ee
for some specific functions $a(\tau)$, $b(\tau)$, and $c(\tau)$. In order to compare equations \eqref{dsy} and \eqref{cglp}, we need the relations
\be
  \sum_i d\by_i dy_i \eq \half \sigma_i^2 + \half \tilde{\sigma}_i^2 + \nu^2 \; , \\
  \bigabs{\sum_i \by_i dy_i}^2 \eq \nu^2 \; , \\
  \sum_i \bigbrk{ dy_i dy_i + d\by_i d\by_i } \eq  \sigma_i^2 - \tilde{\sigma}_i^2 \; .
\ee
Then it follows that
\be
  a^2 \eq \Half (\coth\tau+\csch\tau)\mathcal{F}'
  \; , \\
  b^2 \eq \Half (\coth\tau-\csch\tau)\mathcal{F}'
  \; , \\
  c^2 \eq \mathcal{F}'' = \frac{d-2}{d-1} \lrbrk{ \frac{\epsilon^2 \sinh\tau}{\mathcal{F}'} }^{d-2}
  \; .
\ee

\bibliographystyle{ssg}
\bibliography{glueballs}

\begingroup\raggedright\begin{thebibliography}{10}

\bibitem{Maldacena:1997re}
J.~M. Maldacena, ``{The large $N$ limit of superconformal field theories and
  supergravity},'' {\em Adv. Theor. Math. Phys.} {\bf 2} (1998) 231--252,
  \href{http://xxx.lanl.gov/abs/hep-th/9711200}{{\tt hep-th/9711200}}.

\bibitem{Gubser:1998bc}
S.~S. Gubser, I.~R. Klebanov, and A.~M. Polyakov, ``Gauge theory correlators
  from non-critical string theory,'' {\em Phys. Lett.} {\bf B428} (1998)
  105--114, \href{http://xxx.lanl.gov/abs/hep-th/9802109}{{\tt
  hep-th/9802109}}.

\bibitem{Witten:1998qj}
E.~Witten, ``{Anti-de Sitter space and holography},'' {\em Adv. Theor. Math.
  Phys.} {\bf 2} (1998) 253--291,
  \href{http://xxx.lanl.gov/abs/hep-th/9802150}{{\tt hep-th/9802150}}.

\bibitem{Witten:1998zw}
E.~Witten, ``{Anti-de Sitter space, thermal phase transition, and confinement
  in gauge theories},'' {\em Adv. Theor. Math. Phys.} {\bf 2} (1998) 505--532,
  \href{http://xxx.lanl.gov/abs/hep-th/9803131}{{\tt hep-th/9803131}}.

\bibitem{Klebanov:2000hb}
I.~R. Klebanov and M.~J. Strassler, ``{Supergravity and a confining gauge
  theory: Duality cascades and $\chi_{\text{SB}}$-resolution of naked
  singularities},'' {\em JHEP} {\bf 08} (2000) 052,
  \href{http://xxx.lanl.gov/abs/hep-th/0007191}{{\tt hep-th/0007191}}.

\bibitem{Csaki:1998qr}
C.~Csaki, H.~Ooguri, Y.~Oz, and J.~Terning, ``{Glueball mass spectrum from
  supergravity},'' {\em JHEP} {\bf 01} (1999) 017,
  \href{http://xxx.lanl.gov/abs/hep-th/9806021}{{\tt hep-th/9806021}}.

\bibitem{deMelloKoch:1998qs}
R.~de~Mello~Koch, A.~Jevicki, M.~Mihailescu, and J.~P. Nunes, ``{Evaluation Of
  Glueball Masses From Supergravity},'' {\em Phys. Rev.} {\bf D58} (1998)
  105009, \href{http://xxx.lanl.gov/abs/hep-th/9806125}{{\tt hep-th/9806125}}.

\bibitem{Brower:2000rp}
R.~C. Brower, S.~D. Mathur, and C.-I. Tan, ``{Glueball Spectrum for QCD from
  AdS Supergravity Duality},'' {\em Nucl. Phys.} {\bf B587} (2000) 249--276,
  \href{http://xxx.lanl.gov/abs/hep-th/0003115}{{\tt hep-th/0003115}}.

\bibitem{Ooguri:1998hq}
H.~Ooguri, H.~Robins, and J.~Tannenhauser, ``{Glueballs and their Kaluza-Klein
  cousins},'' {\em Phys. Lett.} {\bf B437} (1998) 77--81,
  \href{http://xxx.lanl.gov/abs/hep-th/9806171}{{\tt hep-th/9806171}}.

\bibitem{Klebanov:2000nc}
I.~R. Klebanov and A.~A. Tseytlin, ``{Gravity Duals of Supersymmetric $SU(N)
  \times SU(N+M)$ Gauge Theories},'' {\em Nucl. Phys.} {\bf B578} (2000)
  123--138, \href{http://xxx.lanl.gov/abs/hep-th/0002159}{{\tt
  hep-th/0002159}}.

\bibitem{Candelas:1989js}
P.~Candelas and X.~C. de~la Ossa, ``{Comments on Conifolds},'' {\em Nucl.
  Phys.} {\bf B342} (1990) 246--268.

\bibitem{Krasnitz:2000ir}
M.~Krasnitz, ``{A two point function in a cascading N = 1 gauge theory from
  supergravity},'' \href{http://xxx.lanl.gov/abs/hep-th/0011179}{{\tt
  hep-th/0011179}}.

\bibitem{Gubser:2004qj}
S.~S. Gubser, C.~P. Herzog, and I.~R. Klebanov, ``{Symmetry breaking and
  axionic strings in the warped deformed conifold},'' {\em JHEP} {\bf 09}
  (2004) 036, \href{http://xxx.lanl.gov/abs/hep-th/0405282}{{\tt
  hep-th/0405282}}.

\bibitem{Berg:2005pd}
M.~Berg, M.~Haack, and W.~Mueck, ``{Bulk dynamics in confining gauge
  theories},'' {\em Nucl. Phys.} {\bf B736} (2006) 82--132,
  \href{http://xxx.lanl.gov/abs/hep-th/0507285}{{\tt hep-th/0507285}}.

\bibitem{Berg:2006xy}
M.~Berg, M.~Haack, and W.~Mueck, ``{Glueballs vs. gluinoballs: Fluctuation
  spectra in non- AdS/non-CFT},'' {\em Nucl. Phys.} {\bf B789} (2008) 1--44,
  \href{http://xxx.lanl.gov/abs/hep-th/0612224}{{\tt hep-th/0612224}}.

\bibitem{Dymarsky:2007zs}
A.~Dymarsky and D.~Melnikov, ``{Gravity Multiplet on KS and BB Backgrounds},''
  {\em JHEP} {\bf 05} (2008) 035, \href{http://xxx.lanl.gov/abs/0710.4517}{{\tt
  0710.4517}}.

\bibitem{Benna:2007mb}
M.~K. Benna, A.~Dymarsky, I.~R. Klebanov, and A.~Solovyov, ``{On Normal Modes
  of a Warped Throat},'' {\em JHEP} {\bf 06} (2008) 070,
  \href{http://xxx.lanl.gov/abs/0712.4404}{{\tt 0712.4404}}.

\bibitem{Dymarsky:2008wd}
A.~Dymarsky, D.~Melnikov, and A.~Solovyov, ``{I-odd sector of the
  Klebanov-Strassler theory},'' {\em JHEP} {\bf 05} (2009) 105,
  \href{http://xxx.lanl.gov/abs/0810.5666}{{\tt 0810.5666}}.

\bibitem{Gordeli:2009nw}
I.~Gordeli and D.~Melnikov, ``{On I-even Singlet Glueballs in the
  Klebanov-Strassler Theory},'' \href{http://xxx.lanl.gov/abs/0912.5517}{{\tt
  0912.5517}}.

\bibitem{Firouzjahi:2005qs}
H.~Firouzjahi and S.~H.~H. Tye, ``{The shape of gravity in a warped deformed
  conifold},'' {\em JHEP} {\bf 01} (2006) 136,
  \href{http://xxx.lanl.gov/abs/hep-th/0512076}{{\tt hep-th/0512076}}.

\bibitem{Herzog:2001xk}
C.~P. Herzog, I.~R. Klebanov, and P.~Ouyang, ``{Remarks on the warped deformed
  conifold},'' \href{http://xxx.lanl.gov/abs/hep-th/0108101}{{\tt
  hep-th/0108101}}.

\bibitem{Klebanov:1998hh}
I.~R. Klebanov and E.~Witten, ``{Superconformal field theory on threebranes at
  a Calabi-Yau singularity},'' {\em Nucl. Phys.} {\bf B536} (1998) 199--218,
  \href{http://xxx.lanl.gov/abs/hep-th/9807080}{{\tt hep-th/9807080}}.

\bibitem{Klebanov:2002gr}
I.~R. Klebanov, P.~Ouyang, and E.~Witten, ``{A gravity dual of the chiral
  anomaly},'' {\em Phys. Rev.} {\bf D65} (2002) 105007,
  \href{http://xxx.lanl.gov/abs/hep-th/0202056}{{\tt hep-th/0202056}}.

\bibitem{Krishnan:2008gx}
C.~Krishnan and S.~Kuperstein, ``{The Mesonic Branch of the Deformed
  Conifold},'' {\em JHEP} {\bf 05} (2008) 072,
  \href{http://xxx.lanl.gov/abs/0802.3674}{{\tt 0802.3674}}.

\bibitem{Dymarsky:2005xt}
A.~Dymarsky, I.~R. Klebanov, and N.~Seiberg, ``{On the moduli space of the
  cascading SU(M+p) x SU(p) gauge theory},'' {\em JHEP} {\bf 01} (2006) 155,
  \href{http://xxx.lanl.gov/abs/hep-th/0511254}{{\tt hep-th/0511254}}.

\bibitem{Klebanov:2007us}
I.~R. Klebanov and A.~Murugan, ``{Gauge/Gravity Duality and Warped Resolved
  Conifold},'' {\em JHEP} {\bf 03} (2007) 042,
  \href{http://xxx.lanl.gov/abs/hep-th/0701064}{{\tt hep-th/0701064}}.

\bibitem{Cvetic:2000db}
M.~Cvetic, G.~W. Gibbons, H.~Lu, and C.~N. Pope, ``{Ricci-flat metrics,
  harmonic forms and brane resolutions},'' {\em Commun. Math. Phys.} {\bf 232}
  (2003) 457--500, \href{http://xxx.lanl.gov/abs/hep-th/0012011}{{\tt
  hep-th/0012011}}.

\bibitem{Stenzel}
M.~Stenzel, ``{Ricci-flat Metrics on the Complexification of a Compact Rank One
  Symmetric Space},'' {\em Manuscripta Mathematica} {\bf 80} (1993) 151.

\bibitem{Herzog:2000rz}
C.~P. Herzog and I.~R. Klebanov, ``{Gravity duals of fractional branes in
  various dimensions},'' {\em Phys. Rev.} {\bf D63} (2001) 126005,
  \href{http://xxx.lanl.gov/abs/hep-th/0101020}{{\tt hep-th/0101020}}.

\bibitem{Martelli:2009ga}
D.~Martelli and J.~Sparks, ``{$AdS_4$/CFT${}_3$ duals from M2-branes at
  hypersurface singularities and their deformations},'' {\em JHEP} {\bf 12}
  (2009) 017, \href{http://xxx.lanl.gov/abs/0909.2036}{{\tt 0909.2036}}.

\bibitem{Jafferis:2009th}
D.~L. Jafferis, ``{Quantum corrections to N=2 Chern-Simons theories with flavor
  and their AdS4 duals},'' \href{http://xxx.lanl.gov/abs/0911.4324}{{\tt
  0911.4324}}.

\bibitem{JLthesis}
J.~Lin, {\em {Bound state spectrum from gauge/gravity duality}}.
\newblock Senior thesis, Princeton University, 2010.

\bibitem{Eguchi:1978xp}
T.~Eguchi and A.~J. Hanson, ``{Asymptotically Flat Selfdual Solutions to
  Euclidean Gravity},'' {\em Phys. Lett.} {\bf B74} (1978) 249.

\bibitem{Kervaire}
M.~A. Kervaire, ``Non-parallelizability of the $n$-sphere for $n>7$,'' {\em
  Proc. Nat. Acad. Sci.} {\bf 44} (1958) 280--283.

\bibitem{MR0102804}
R.~Bott and J.~Milnor, ``On the parallelizability of the spheres,'' {\em Bull.
  Amer. Math. Soc.} {\bf 64} (1958) 87--89.

\bibitem{Hatcher}
A.~Hatcher, {\em {Algebraic topology}}.
\newblock Cambridge University Press, 2002.

\bibitem{Strichartz}
R.~S. Strichartz, ``{The Explicit Fourier decomposition of
  $L^2(\grSO(n)/\grSO(n-m)$},'' {\em Can. J. Math.} {\bf 27} (1975) 294.

\bibitem{Gelbart1}
S.~S. Gelbart, ``{Harmonics on Stiefel manifolds and generalized Hankel
  transforms},'' {\em Bull. Amer. Math. Soc.} {\bf 78} (1972) 451.

\bibitem{Gelbart2}
S.~S. Gelbart, ``{A theory of Stiefel harmonics},'' {\em Trans. Amer. Math.
  Soc.} {\bf 192} (1974) 29.

\bibitem{Levine}
D.~A. Levine, ``{Systems of singular integral operators on spheres},'' {\em
  Trans. Amer. Math. Soc.} {\bf 144} (1969) 493.

\bibitem{Taylor}
M.~E. Taylor, {\em {Noncommutative Harmonic Analysis}}.
\newblock American Mathematical Society, 1986.

\bibitem{Romans:1984an}
L.~J. Romans, ``{New compactifications of chiral N=2 d = 10 supergravity},''
  {\em Phys. Lett.} {\bf B153} (1985) 392.

\bibitem{Ceresole:1999zg}
A.~Ceresole, G.~Dall'Agata, R.~D'Auria, and S.~Ferrara, ``{M-theory on the
  Stiefel manifold and 3d conformal field theories},'' {\em JHEP} {\bf 03}
  (2000) 011, \href{http://xxx.lanl.gov/abs/hep-th/9912107}{{\tt
  hep-th/9912107}}.

\bibitem{Gubser:1998vd}
S.~S. Gubser, ``{Einstein manifolds and conformal field theories},'' {\em Phys.
  Rev.} {\bf D59} (1999) 025006,
  \href{http://xxx.lanl.gov/abs/hep-th/9807164}{{\tt hep-th/9807164}}.

\bibitem{Ceresole:1999zs}
A.~Ceresole, G.~Dall'Agata, R.~D'Auria, and S.~Ferrara, ``{Spectrum of type IIB
  supergravity on $AdS_5 \times T^{1, 1}$: Predictions on ${\cal N} = 1$
  SCFT's},'' {\em Phys. Rev.} {\bf D61} (2000) 066001,
  \href{http://xxx.lanl.gov/abs/hep-th/9905226}{{\tt hep-th/9905226}}.

\bibitem{GlueballWebsite}
{\tt home.uchicago.edu/$\sim$jenlin/Glueballs/}.

\bibitem{Page:1979ga}
D.~N. Page, ``{Green's functions for gravitational multi-instantons},'' {\em
  Phys. Lett.} {\bf B85} (1979) 369.

\bibitem{Girardi:1981jz}
G.~Girardi, A.~Sciarrino, and P.~Sorba, ``{Kronecker products for SO(2p)
  representations},'' {\em J. Phys.} {\bf A15} (1982) 1119.

\bibitem{Girardi:1981ss}
G.~Girardi, A.~Sciarrino, and P.~Sorba, ``{Generalized Young tableaux and
  Kronecker products of SO(n) representations},'' {\em Physica} {\bf 114A}
  (1982) 365.

\bibitem{Klebanov:2010qs}
I.~R. Klebanov and S.~S. Pufu, ``{M-Branes and Metastable States},''
  \href{http://xxx.lanl.gov/abs/1006.3587}{{\tt 1006.3587}}.

\bibitem{Kachru:2002gs}
S.~Kachru, J.~Pearson, and H.~L. Verlinde, ``{Brane/Flux Annihilation and the
  String Dual of a Non- Supersymmetric Field Theory},'' {\em JHEP} {\bf 06}
  (2002) 021, \href{http://xxx.lanl.gov/abs/hep-th/0112197}{{\tt
  hep-th/0112197}}.

\bibitem{Bena:2009xk}
I.~Bena, M.~Grana, and N.~Halmagyi, ``{On the Existence of Meta-stable Vacua in
  Klebanov- Strassler},'' \href{http://xxx.lanl.gov/abs/0912.3519}{{\tt
  0912.3519}}.

\end{thebibliography}\endgroup

\end{document}